\documentclass[11pt,tightenlines,eqsecnum,floats,aps,amsmath,amssymb,nofootinbib,prd,shownopacs]{revtex4}
\usepackage{amsmath,amssymb,amsfonts,amsthm}
\usepackage{graphicx}
\usepackage{enumerate} % advanced enumerate environment
\usepackage{colordvi} % for color text
\usepackage[mathscr]{eucal}
\usepackage[latin1]{inputenc}
\usepackage{amsmath}
\usepackage{amsfonts}
\usepackage{amssymb}
\usepackage{graphicx}

\def\lp{{\ell}_{\rm Pl}}

\def\q{\mathring{q}}
\def\e{\mathring{e}}
\def\w{\mathring{\omega}}
\def\ow{{}^o\!\omega}

\newcommand{\rcr}{\rho_{\mathrm{crit}}}

\newcommand{\p}{\partial}

\newcommand{\f}{\frac}

\def\rmax{\rho_{\mathrm{max}}}

\def\rcr{\rmax}

\usepackage{enumerate}

\def\f{\frac}

\def\d{\textrm{d}}

\usepackage{enumerate}

\newcommand{\be}{\nopagebreak[3]\begin{equation}}
\newcommand{\ee}{\end{equation}}
\newcommand{\ba}{\nopagebreak[3]\begin{eqnarray}}
\newcommand{\ea}{\end{eqnarray}}

\newcommand{\bmult}{\nopagebreak[3]\begin{multline}}
\newcommand{\emult}{\end{multline}}

\def\lp{{\ell}_{\rm Pl}}

\def\ow{{}^o\!\omega}

\def\f{\frac}

\def\d{\textrm{d}}

\def\rmax{\rho_{\rm max}}

\def\lp{l_{\rm Pl}}

\def\d{{\rm d}}
\usepackage{enumerate}
\begin{document}

\title{Non-singular Power-law and Assisted inflation in Loop Quantum Cosmology}

\author{Evan Ranken}
\email{evan.ranken@coloradocollege.edu}
%
%Baton Rouge, Louisiana 70803, USA}

\author{Parampreet Singh}
\email{psingh@phys.lsu.edu}
%\affiliation{}
\affiliation{${}^*$Department of Physics, Colorado College, Colorado Springs, Colorado 80903, USA \\
${}^\dagger$Department of Physics and Astronomy, Louisiana State University,
Baton Rouge, Louisiana 70803, USA}


\begin{abstract}
We investigate the dynamics of single and multiple scalar fields with exponential potentials, leading to power-law and assisted inflation, in loop quantum cosmology. Unlike in the classical theory, dynamical trajectories in loop quantum cosmology are generically non-singular, with a big bounce replacing classical big bang in the Planck regime. Post bounce, after a phase of super-inflation, dynamical trajectories evolve towards the classical attractor in the inflationary scenarios. Depending on the initial conditions, bounce is shown to occur in kinetic as well as potential dominated regimes.
%Phase space analysis reveals attractors in the pre-bounce and post-bounce phases, which approximate the classical attractors at very early and late times respectively.
We analyze the number of e-foldings resulting from the phase of super-inflation, and find the dependence of the maximum possible number of e-foldings on the equation of state at the bounce and on the steepness of the potential. We find that if the potential is not steep, this phase can lead to large number of e-foldings in power-law inflation. For the assisted inflation scenario, an increase in the number of fields can yield a significant increase in the number of e-foldings during super-inflation.
\end{abstract}

%%%%%%%%%%%%%%%%%%%%%%%%
%%%%%%%%%%%%%%%%%%%%%%%%
\maketitle
%%%%%%%%%%%%%%%%%%%%%%%%
%%%%%%%%%%%%%%%%%%%%%%%%

%***********************************************************************

\section{Introduction}
The inflationary paradigm provides one of the simplest ways to describe various aspects of the physics of the early universe in standard cosmology. However, despite its successes some of the fundamental questions yet remain  to be answered in this paradigm. These include the problem of initial singularity and the past incompleteness of inflationary spacetimes, solution of which are expected to be provided by a quantum theory of gravity. Another issue is to understand the nature of inflaton itself. Though a simple $\phi^2$ inflaton potential fits extremely well with the observations, the origin of such a potential from  high energy physics remains unclear. Theories invoking higher dimensions and supersymmetry yield multiple fields in the effective 4-dimensional description of the early universe. Further, models of supergravity naturally lead to exponential potentials \cite{higherdim}, which do not yield exponential inflation as in the $\phi^2$ potential. Such potentials lead to a power-law inflation provided the potential is not too steep \cite{lucchin}. The limitation of steep potentials can be alleviated using multiple fields, which can cooperate via Hubble damping to yield assisted inflation \cite{assisted1}. Interestingly, dynamical evolution in assisted inflation can also be understood as an effective single field power-law inflation.

The dynamical properties of the positive exponential potentials leading to inflation in the Friedmann-Robertson-Walker (FRW) model have been widely studied \cite{halliwell,barrow,attrac1,coley1,coley2,assisted2,assisted3,assisted_cross,green_lidsey,assisted_b1,attrac2,piao,wands}.\footnote{Negative exponential potentials also lead to a rich physics, such as in Ekpyrotic scenario \cite{ekpy}. For an analysis of such a potential on the lines of the present work in loop quantum cosmology, we refer the reader to Refs. \cite{svv,csv}.}
In the forward evolution in the expanding branch, for sufficiently flat potentials, exponential potentials yield a scaling solution which is a late time attractor \cite{halliwell,barrow,attrac1}. However, in the past evolution, dynamical trajectories in the power-law and assisted inflation scenarios lead to a kinetic energy  dominated regime as the big bang singularity at vanishing scale factor is approached. Higher dimensional and supergravity models which yield these  exponential potentials do not provide a generic remedy to overcome this fundamental problem. It has been hoped that incorporation of non-perturbative quantum gravitational effects would lead to the resolution of the singularity problem. New physics resulting from such a non-perturbative quantization can also potentially affect the inflationary paradigm itself, and in principle may alleviate the observational constraints on power-law and assisted inflationary models.

In recent years,  quantization of homogeneous models in loop quantum cosmology (LQC) have shown that the cosmological singularities in various models can be successfully resolved \cite{as}. LQC is a canonical quantization of cosmological spacetimes based on loop quantum gravity (LQG), a background independent and non-perturbative approach to quantize gravity. A key prediction of LQG is that the continuous differential geometry of the classical theory is replaced by a discrete quantum geometry at the Planck scale. The discreteness of quantum geometry has various novel implications. The Hamiltonian constraint, whose vanishing  determines the physical solutions, is a quantum difference operator in LQC, and the resulting evolution equation is non-singular with uniform steps in the physical volume of the universe. The step size of the difference equation is determined by the minimum allowed eigenvalue of the area operator in LQG. At small spacetime curvature, the quantum difference equation can be approximated as a differential equation, and the solutions of the Hamiltonian constraint in LQC approach those of the Wheeler-DeWitt theory. In contrast, significant departures between LQC and Wheeler-DeWitt theory occur when spacetime curvature approaches Planck regime. A state peaked on the classical trajectory when evolved backwards using quantum Hamiltonian constraint does not encounter a big bang singularity, but bounces when the energy density reaches a critical value $\rho = \rmax \approx 0.41 \rho_{\rm{Pl}}$ (where $\rho_{\rm{Pl}}$ denotes Planck density) in the isotropic models \cite{aps1,aps2,aps3}.\footnote{The occurrence of bounce at $\rmax$ is only a property of isotropic cases in LQC. In anisotropic models, bounces of the anisotropic scale factors are characterized by the behavior of energy density of the matter field and the shear scalar \cite{gs1}. Bounces in anisotropic models in LQC can occur even when energy density and shear scalar do not attain a maximum.} The existence of the bounce has been demonstrated to be a robust feature of the theory \cite{acs}. The singularity resolution at the quantum level occurs independent of the initial conditions or the choice of parameters for various matter models, including models with a non-vanishing spatial curvature \cite{apsv,kv,szulc}, cosmological constant \cite{bp,ap}, massive scalar field \cite{aps3} and anisotropic models \cite{awe2,awe3,we1}. Using consistent histories approach, it has also been shown that the probability that a loop quantum universe encounters a singularity is zero \cite{cs3}.\footnote{In contrast to the result in the Wheeler-DeWitt theory, where the answer turns out to be unity \cite{cs1,cs2}.}

For states which lead to a macroscopic universe at late times, discrete quantum evolution in LQC can be approximated by an effective continuum spacetime description using geometric formulation of quantum mechanics \cite{josh,vt,psvt}. Effective dynamics has been widely used in LQC to understand the physical implications in various models and has provided important insights on the genericity of singularity resolution in LQC. In particular, analysis of isotropic and anisotropic models indicate that all strong curvature singularities in LQC are resolved \cite{ps09,sv,ps11}. Using effective dynamics, effects of the loop quantum geometric corrections on $\phi^2$ inflationary model have also been studied \cite{infl-lqc1,infl-lqc2,infl-lqc3,ps06,svv,infl-lqc4,infl-lqc5,infl-lqc6,as_infl,ck}. The phase of inflation turns out to be an attractor of dynamical trajectories after bounce \cite{svv} and is preceded by a phase of super-inflation, where the rate of change of the Hubble rate is positive, for all matter obeying the weak energy condition \cite{ps06}. These properties have also been used to understand the likelihood of the occurrence of inflation for the $\phi^2$ potential \cite{infl-lqc6,as_infl,ck}.

 In this work we analyze the dynamics of exponential potential leading to power-law and assisted inflation scenarios. We show that independent of the choice of initial conditions, effective dynamics is devoid of any singularities and power-law and assisted inflation fits naturally in LQC. In LQC, the scale factor bounces when energy density becomes equal to $\rmax$, leading to a phase of super-inflation which exits to the phase of classical inflation or decelerated expansion, depending on the steepness of the potential. Analysis of the phase space trajectories in $\phi-\dot \phi$ plane shows that in the regime $\rho \lesssim \rmax/2$, dynamical trajectories in LQC converge towards a curve which approaches the late time attractor in the classical theory in the forward time evolution when $\rho \ll \rmax$. In the backward time evolution, dynamical trajectories in the pre-bounce regime, approach the classical attractor at very early times in the contracting branch. In the classical theory, since contracting and expanding branches are disjoint, classical solutions either approach the forward time attractor (in the expanding branch) or are repelled by the forward time repeller (in the contracting branch).
In LQC, quantum geometric effects lead to a non-singular evolution between the above attractor and repeller. In the classical theory, in the backward evolution of the expanding branch, dynamical trajectories approach those of the massless scalar near the singularity (in agreement with the Foster's theorem \cite{foster}).  Hence, classical  dynamics is governed by kinetic energy of the scalar field near the big bang/crunch with the equation of state $w$ approaching unity. In contrast, in LQC, we find that the equation of state with the positive exponential potential can range from $w = 1$ to $w=-1$ at the bounce, depending on the initial conditions. %In the classical theory, under mild assumptions, all positive potentials share this feature as proved by Foster's theorem in classical theory \cite{foster}.
 In LQC, though Foster's theorem is not applicable, we find that it is easier to find trajectories with a  bounce where kinetic energy dominates over potential energy, i.e. with equation of state  $w \sim 1$ in the backward evolution of the expanding branch.

 Since LQC leads to a phase of super-inflation preceding the inflationary phase, there are e-foldings which have a pure quantum gravitational origin.
In the $\phi^2$ potential, unless at the bounce equation of state is $w \sim -1$, the number of e-foldings in the super-inflationary regime turn out to be very small \cite{infl-lqc5,as_infl}. A pertinent question is what specific conditions on the exponential potentials will lead to a significant number of e-foldings during super-inflation and whether these potentials can further assist in the assisted inflation scenario.  We find that for the exponential potential, the number of these e-foldings depends in a non-linear way on the equation of state at the bounce and the steepness of potential. As the steepness of the potential decreases, the maximum possible number of e-foldings during super-inflation occur for $w \rightarrow -1$. However, for a non-zero value of the steepness parameter for the exponential potential, the maximum number of e-foldings in super-inflation do not occur for $w=-1$. Further, number of e-foldings can be significant even if $w$ is not approximately $-1$ at the bounce. We analyze the issue of number of e-foldings during super-inflation for multiple field scenario and find that the number of e-foldings during super-inflation almost always turns out to be smaller than the equivalent single field scenario. Our investigations also reveal that by increasing the number of fields in assisted inflation scenario, the number of e-foldings during super-inflation phase can increase.

In the following we begin with an overview of derivation of classical Friedmann and Raychaudhuri equations using the connection and triad variables of loop quantum gravity. We show the way these equations can be obtained in the canonical framework. These are then used to derive the modified Friedmann and Raychaudhuri equations using effective Hamiltonian in Sec. IIB. In Sec. III, we use the effective dynamical equations to investigate the case of power-law inflation. A detailed analysis of the behavior of dynamical trajectories and super-inflationary phase is presented, along with the dependence of number of e-foldings during super-inflation on steepness parameter of exponential potential and the equation of state at the bounce.  This is followed by the analysis of multiple fields in Sec. IV leading to assisted inflation. Since system with multiple fields can be expressed as an equivalent single field case in assisted inflation, various results obtained in Sec. III immediately apply to the cases in Sec. IV. We conclude with a summary of results in Sec. V.

%In the quantum theory, there exists no corresponding operator for the connection. Instead the elementary variable is the holonomy of connection over a loop, with a minimum allowed area on the underlying quantum geometry, given by LQG. The underlying quantum discreteness manifests itself in the quantum Hamiltonian constraint which
%turns out to be a difference equation with uniform steps in volume, with a step size determined by the minimum area of the loop over which holonomy is computed \cite{aps3}.  Unlike in Wheeler-DeWitt theory, evolution in LQC turns out to be non-singular, with a quantum big bounce replacing classical big bang for various matter models \cite{as}.

\section{Classical and effective dynamics for $k=0$ isotropic model}
In this section, we review the way dynamical equations in the classical theory and the effective spacetime description of LQC can be derived in the Hamiltonian framework. The spacetime metric for the spatially flat model with lapse $N$ chosen as unity is given by
\be
\d s^2 = - \d t^2 + a^2(t) (\d x_1^2 + \d x_2^2 + \d x_3^2) ~,
\ee
where $a(t)$ is the scale factor of the universe. We will consider the spatial manifold to be $\mathbb{R}^3$ with a
physical metric $q_{ab}$. Since the spatial manifold is non-compact, in order to define  symplectic structure we introduce a fiducial cell ${\cal C}$, which plays the role of an infra-red regulator to control divergences in the spatial integrals.\footnote{The choice of the fiducial cell must not affect the physical predictions resulting from the quantization. For a discussion of these issues in LQC, and the way unphysical quantizations can be ruled out, we refer the reader to Ref. \cite{cs08,cs09}.} The physical volume of ${\cal C}$ is $V = a^3 V_o$, where $V_o$ is the volume of the fiducial cell with respect to the fiducial metric $\q_{ab}$, given by $\q_{ab} = a^{-2} q_{ab}$.

The gravitational sector of the classical phase space in LQG is spanned by the Ashtekar-Barbero SU(2) connection $A^i_a$ and the conjugate triad $E^a_i$. In LQC, due to the underlying symmetries of the isotropic spacetime, these variables can be symmetry reduced as
\be\label{AE_defs}
A^i_a \, = \, c  \, V_o^{-1/3} \w^i_a, ~~~
E^a_i \, = \,  p \, \sqrt{\q} \, \,  V_o^{-2/3} \, \e^a_i ~,
\ee
where $\e^a_i$ and $\ow^i_a$ are the fiducial triad and co-triads associated with the cartesian coordinates. The triad $p$ satisfies a kinematical relation, $|p| = V_o^{2/3} a^2$, where the modulus signs arises due to two possible orientations of the triad. In the following, without any loss of generality, we choose the fiducial volume of the cell ${\cal C}$ to be unity and fix the orientation of the triad to be positive. The triad obeys the following Poisson bracket relation with connection $c$:
%The gravitational phase space variables, $c$ and $p$ connection  satisfy
\be
\{c, p\} = \frac{8\pi\gamma G}{3} ~.
\ee
Here $\gamma$ is the Barbero-Immirzi parameter, with value $\gamma \approx 0.2375$ fixed by the black hole thermodynamics in LQG. In the following, we first discuss the classical Hamiltonian constraint and the derivation of Friedmann and Raychaudhuri equations in the canonical framework. This is followed by a similar derivation for the effective dynamics in LQC.

\subsection{Classical dynamics in Ashtekar variables}
Dynamical equations of the classical theory can be derived starting from the Hamiltonian constraint expressed in terms of $c,p$ variables and using Hamilton's equations. The classical Hamiltonian constraint is given by,
\be\label{clH}
{\cal H}_{\mathrm{cl}} = - \f{3}{8 \pi G \gamma^2} \, p^{1/2} \, c^2 + {\cal H}_{\mathrm{matt}} \approx 0
\ee
where ${\cal H}_{\mathrm{matt}}$ corresponds to the matter Hamiltonian, which in the  case of minimally coupled scalar fields $\phi_i$, each with a potential $V_i(\phi_i)$, is given by
\be\label{clHm}
{\cal H}_{\mathrm{matt}} = \sum_i {\cal H}_{\mathrm{matt}(i)} = \sum_i \left(\f{1}{2} \f{p_{\phi_i}^2}{p^{3/2}} + p^{3/2} V_i(\phi_i)\right) ~.
\ee
The energy density $\rho_i$ and pressure $P_i$ of a field $\phi_i$, are defined as $\rho_i = {\cal H}_{\mathrm{matt}(i)}/p^{3/2}$ and $P_i = - \p {\cal H}_{\mathrm{matt}(i)}/\p p^{3/2}$, which turn out to be,
\be\label{clden}
\rho_i = \f{1}{2}  \f{p_{\phi_i}^2}{p^{3}} + V_i(\phi_i) , ~~~ P_i = \f{1}{2}  \f{p_{\phi_i}^2}{p^{3}} - V_i(\phi_i) ~.
\ee
Using the Hamilton's equations for matter variables we obtain,
\be \label{hamilton_matter}
\dot \phi_i = \{\phi_i,{\cal H}_{\mathrm{cl}}\} = \f{p_{\phi_i}}{p^{3/2}} ~, ~~~ \dot p_{\phi_i} = \{p_{\phi_i},{\cal H}_{\mathrm{cl}}\} = - p^{3/2} \, \p_{\phi_i} V_i(\phi_i) ~.
\ee
The time derivative of $\dot \phi_i$, then yields the Klein-Gordon equation for the field $\phi_i$:
\be\label{kg}
\ddot \phi_i + 3 H \dot \phi_i + \p_{\phi_i} V_i(\phi_i) = 0
\ee
where $H$ denotes the Hubble rate, $H = \dot a/a$. Using eqs.(\ref{hamilton_matter}) in expressions of $\rho_i$ and $P_i$ and using (\ref{kg}), it is straightforward to deduce that the total energy density $\rho$ and pressure $P$ satisfy the
conservation law
\be\label{cons}
\dot \rho + 3 H (\rho + P) = 0
\ee
where,
\be
\rho = \sum_i \rho_i = \sum_i \left(\f{1}{2} \dot \phi_i^2 + V(\phi_i)\right) ~~ \mathrm{and} ~~ P = \sum_i P_i = \sum_i \left(\f{1}{2} \dot \phi_i^2 - V(\phi_i)\right) ~.
\ee

We now obtain the Friedmann and Raychaudhuri equations using the Hamilton's equations for the gravitational variables $c$ and $p$. Hamilton's equation for the triad yields,
\be\label{pdot}
\dot p = \{p,{\cal H}_{\mathrm{cl}}\} = - \f{8 \pi G \gamma}{3} \f{\p}{\p c} {\cal H}_{\mathrm{cl}} = \f{2 c}{\gamma} p^{1/2},
\ee
which using $p = a^2$ leads to the relationship between the connection $c$ and the time derivative of the scale factor: $c = \gamma \dot a$. Similarly, Hamilton's equation for $c$ leads to,
\be\label{cdot}
\dot c = \{c,{\cal H}_{\mathrm{cl}}\} = \f{8 \pi G \gamma}{3} \f{\p}{\p p} {\cal H}_{\mathrm{cl}} = - \f{\gamma}{2} H a - 4 \pi G \gamma \, a P
\ee
where in the last step we have used the relation $P = -\p {\cal H}_{\mathrm{matt}}/\p p^{3/2}$ and the eq.(\ref{pdot}). Using the latter, the Friedmann equation follows directly by imposing the Hamiltonian constraint (\ref{clH}), and dividing it by the physical volume,
\be\label{fried}
\f{c^2}{\gamma^2 p} = \f{8 \pi G}{3} \f{\cal H_{\mathrm{matt}}}{p^{3/2}} ~~~ \Longrightarrow ~~~H^2 = \f{8 \pi G}{3} \rho ~.
\ee
Similarly, using  (\ref{cdot}), we get the Raychaudhuri equation:
\be\label{rai}
\f{\ddot a}{a} = \f{1}{\gamma} \f{\dot c}{a} = - \f{4 \pi G}{3} (\rho + 3 P)
\ee
where in the last step we have used the expression of Hubble rate from (\ref{fried}). It is easy to verify that the Friedmann and Raychaudhuri equations together imply the conservation law (\ref{cons}). Thus, we obtain a closed set of dynamical equations for gravitational and matter variables. For the case of power-law and assisted inflationary scenarios, these equations break down as big bang the singularity at $a \rightarrow 0$ is approached. In the regime close to the singularity, kinetic energy of the scalar field dominates the potential energy and solutions behave as in the case of a massless scalar field or equivalently the stiff fluid with equation of state $w = P/\rho = 1$. This behavior is in agreement with the Foster's theorem which shows that for positive potentials as considered here, the asymptotic behavior near the singularity mimics the case of a massless scalar \cite{foster}.   As we will discuss below, the situation completely changes in LQC, where the big bang singularity is generically resolved due to underlying quantum geometric effects.

%A similar set of equations, though incorporating s from the quantum theory, can be derived using the effective Hamiltonian constraint. In LQG, there exists no analog of the connection operator, and the elementary variable is  As we will discuss below, the effective Hamiltonian, due to the use of holonomies,  It turns out that for the spatially flat models, this is the key difference in the Hamiltonian constraints of the classical theory and the effective spacetime description of LQC.

\subsection{Effective dynamics in LQC}
At the quantum level, evolution in LQC is determined by a quantum difference equation with a uniform step in volume. Using geometric formulation of quantum theory, where one treats the Hilbert space as a quantum phase space, it is possible to
derive an effective Hamiltonian for coherent states under controlled approximations \cite{josh,vt,psvt}. Extensive numerical simulations have shown that the dynamical trajectories resulting from effective Hamiltonian turn out to be
in excellent agreement with the underlying quantum evolution for various matter models \cite{aps2,aps3,apsv,kv,bp,ap,aps4,madrid_b1}.
These effective equations incorporate underlying non-perturbative quantum geometric effect in a continuum spacetime description. In comparison to the classical Hamiltonian constraint, the key difference occurs in the gravitational part of the constraint where, instead of $c^2$, the effective Hamiltonian constraint contains trigonometric functions of connection. %This difference can be understood if we recall that the elementary variable in loop quantization is
%the holonomy of the connection which generates an algebra of almost periodic functions with elements of the form $\exp(i \bar \mu c)$.
For the spatially flat isotropic model, the effective Hamiltonian constraint is given by
\be\label{effham}
{H}_{\mathrm{eff}} = H_{\rm{grav}} + H_{\rm{matt}} = - \f{3}{8 \pi G \gamma^2 \bar \mu^2} \, |p|^{1/2} \, \sin^2(\bar \mu c) + {H}_{\mathrm{matt}} \approx 0
\ee
where ${H}_{\mathrm{matt}}$ is considered to be arising from a Fock quantization of the matter Hamiltonian (\ref{clHm}), and $\bar \mu = \lambda/p^{1/2}$. Here $\lambda^2 = 4 (3 \pi \gamma) \lp^2$, with $\lp$ as the Planck length, determined by the quantum geometry \cite{awe2}.  Since the quantum geometric effects do not modify the matter Hamiltonian, Hamilton's equations for matter yield  the same dynamical equations as (\ref{hamilton_matter}). Hence, for the above effective Hamiltonian constraint in LQC, the scalar field satisfies the Klein-Gordon equation (\ref{kg}) and the conservation law (\ref{cons}).

Unlike the dynamical equations for the matter variables, dynamical equations for the gravitational variables turn out to be qualitatively different. Hamilton's equations for the triad yield,
\be\label{dotp_lqc}
\dot p = \{p,{H}_{\mathrm{eff}}\} = - \f{8 \pi G \gamma}{3} \, \f{\p}{\p c} {\cal H}_{\mathrm{eff}} = \f{p^{1/2}}{\gamma \bar \mu} \, \sin(2 \bar \mu c) ~.
\ee
Using this equation, with the vanishing of the effective Hamiltonian constraint
\be
\sin^2(\bar \mu c) = \f{8 \pi G \lambda^2}{3} \, \rho ~,
\ee
it is easy to derive the modified Friedmann equation:
\be\label{modfried}
H^2 = \f{\dot p^2}{4 p^2} = \f{8 \pi G}{3} \, \rho \left(1 - \f{\rho}{\rcr}\right) ~, ~~~ \rm{with} ~~~ \rcr = \f{3}{8 \pi G \gamma^2 \lambda^2} ~.
\ee
Similarly, the time derivative of connection is given by
\be
\dot c = \{c,{H}_{\mathrm{eff}}\} =  \f{8 \pi G \gamma}{3} \, \f{\p}{\p p} {\cal H}_{\mathrm{eff}} = -\f{1}{2\gamma \lambda} \sin(2\bar \mu c)\, c - \f{3}{2 \gamma \lambda^2} \sin^2(\bar \mu c) p^{1/2} - 4 \pi G \gamma a P ~.
\ee
It is then straightforward to obtain the modified Raychaudhuri equation by using the above equation
with the time derivative of (\ref{dotp_lqc}):
\be\label{modrai}
\f{\ddot a}{a} = - \f{4 \pi G}{3} \rho \left(1 - 4 \f{\rho}{\rcr}\right) - 4 \pi G P \, \left(1 - 2 \f{\rho}{\rcr}\right) ~.
\ee
Combining modified Friedmann and Raychaudhuri equations, one obtains another useful equation, measuring the time derivative of the Hubble rate
\be
{\dot H} = - 4 \pi G (\rho + P) \left(1 - 2 \f{\rho}{\rcr}\right) ~.
\ee
As expected from the dynamical equations of the matter variables, the modified Friedmann and Raychaudhuri equations together imply the conservation law (\ref{cons}), as in the classical theory.

We now note some important features of the modified dynamics in LQC. Unlike the classical theory, where energy density can grow without any bound, $\rho$ in LQC is bounded above by a maximum value $\rcr$ which is given by $\rcr \approx 0.41 \rho_{\mathrm{Pl}}$, where $\rho_{\mathrm{Pl}}$ is the Planck density. The Hubble rate attains a maximum value, $|H|_{\mathrm{max}} = 1/(2\gamma^{1/2}\lambda)$ which is reached at $\rho = \rcr/2$. Note that both $\rcr$ and $|H|_{\mathrm{max}}$ diverge as $\lambda \rightarrow 0$. This implies that the classical singularities are recovered in the limit where $\lambda$, the parameter which captures the underlying quantum discreteness in LQC, vanishes. The boundedness of the Hubble rate has novel implications. First, it can be shown that it directly leads to the resolution of all strong singularities in spatially flat isotropic LQC \cite{ps09} (see also Refs. \cite{sv,ps11} for the inclusion of spatial curvature and anisotropy). Secondly, there exists a phase of super-inflation, i.e. $\dot H > 0$, when $\rmax < \rho < \rmax/2$ \cite{ps06}. In literature, many interesting physical implications of this phase have been studied (see Ref. \cite{as} for a review), including the qualitative dynamics of different potentials \cite{svv} and computation of probability for inflation in $\phi^2$ potential to occur \cite{as_infl,ck}.\footnote{Dynamics of bounce and inflation  has also been studied in models not incorporating LQC.  See for eg. \cite{piao1}.} In the following, we will show the way this phase plays an important role in the qualitative dynamics for exponential potentials, which lead to power-law and assisted inflation scenarios.

\section{Exponential potential with a scalar field}
In this section, we analyze the effective dynamics with an exponential potential. Such potentials arise naturally in various higher dimensional frameworks, such as in Kaluza-Klein theories and supergravity \cite{higherdim}, and are of the form,
\be\label{pot1}
V(\phi) = V_o e^{-\sqrt{8 \pi G} \,k \phi}
\ee
where $V_o$ and $k$ are constants. The exponential potential has various interesting properties in GR which have been widely studied in literature (see for eg. \cite{lucchin,halliwell,barrow,attrac1,coley1}) . The classical solution for the scale factor can be written as a power law \cite{lucchin,halliwell,barrow}:
%Using above potential in eqs.(\ref{fried}) and (\ref{rai}), and solving for $a(t)$, one finds that the scale factor has a power-law dependence
\be
a \propto t^n, ~~~ \mathrm{with} ~~~ n = \f{2}{k^2} ~.
\ee
Solving for the  scalar field using classical equations, we obtain  $\phi \propto (2/k) \ln(t)$. The second time derivative of the scale factor implies that the power-law solution is inflationary in the classical theory if $n > 1$. It is non-inflationary if $n < 1$. Thus, in order to have an inflationary phase with the exponential potential, one requires the steepness parameter $k$ to satisfy $k^2 < 2$. If the potential (\ref{pot1}) is steep, such that $k^2 > 2$, then it does not support inflation. Qualitative analysis with potential (\ref{pot1}) shows that near the classical singularity, solutions are kinetic dominated and scale factor behaves as in the case of a massless scalar, i.e. $a \propto t^{1/3}$ \cite{halliwell,attrac2}, as expected from the Foster's theorem \cite{foster}. Further, for steepness parameter such that $k^2 < 6$, which constitute most interesting cases, there exists a potential-kinetic scaling solution which is a late time attractor \cite{attrac2}.\footnote{These properties also generalize to the case of the multi-field potential with $n = $ $\displaystyle\sum\limits_i$ $2/k_i^2$ (see for eg. \cite{coley2,assisted2,piao}).}

\begin{figure}[tbh!]
%\begin{center}
%\label{fig1}
\includegraphics[angle=0,width=0.45\textwidth]{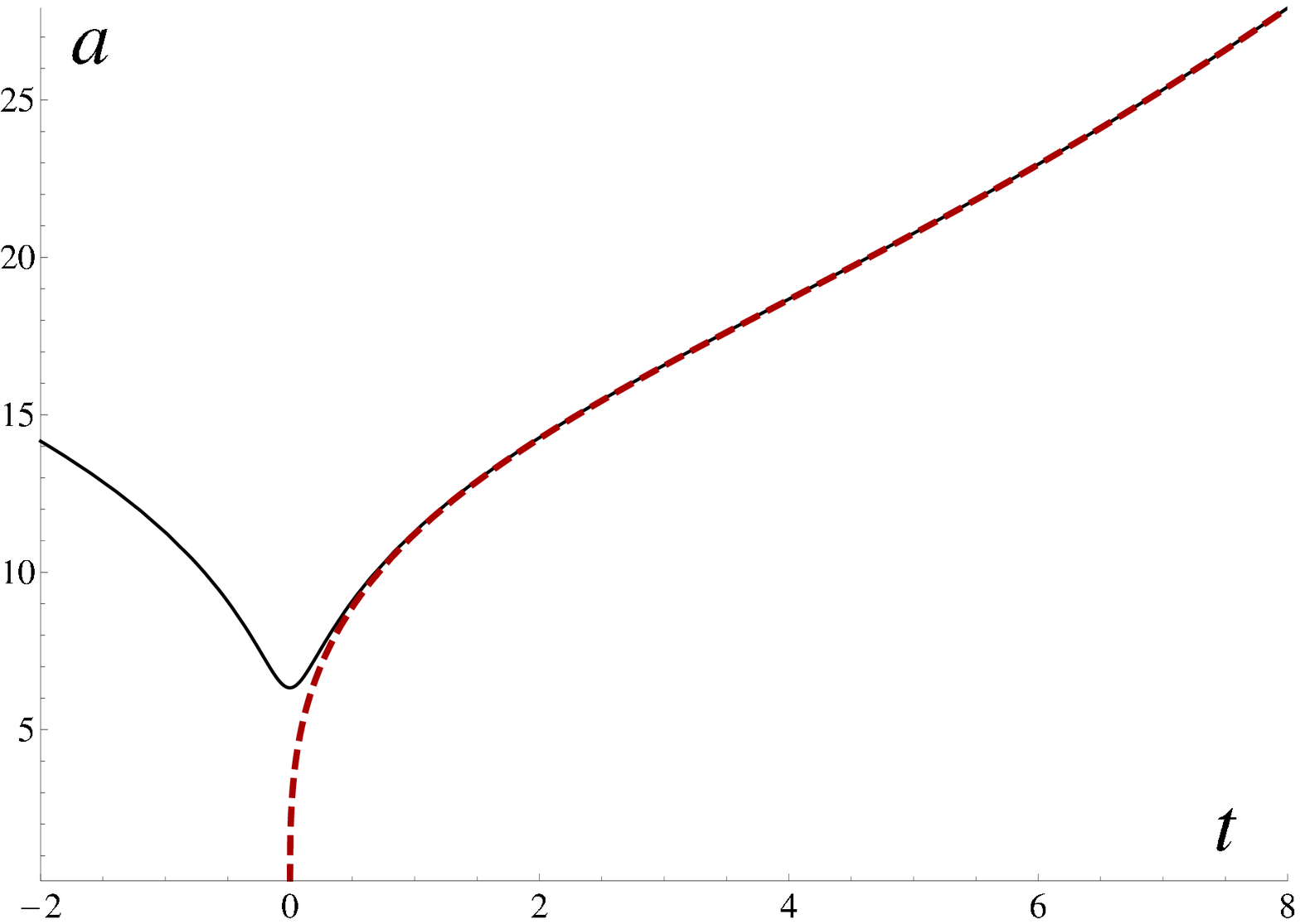}
\includegraphics[angle=0,width=0.45\textwidth]{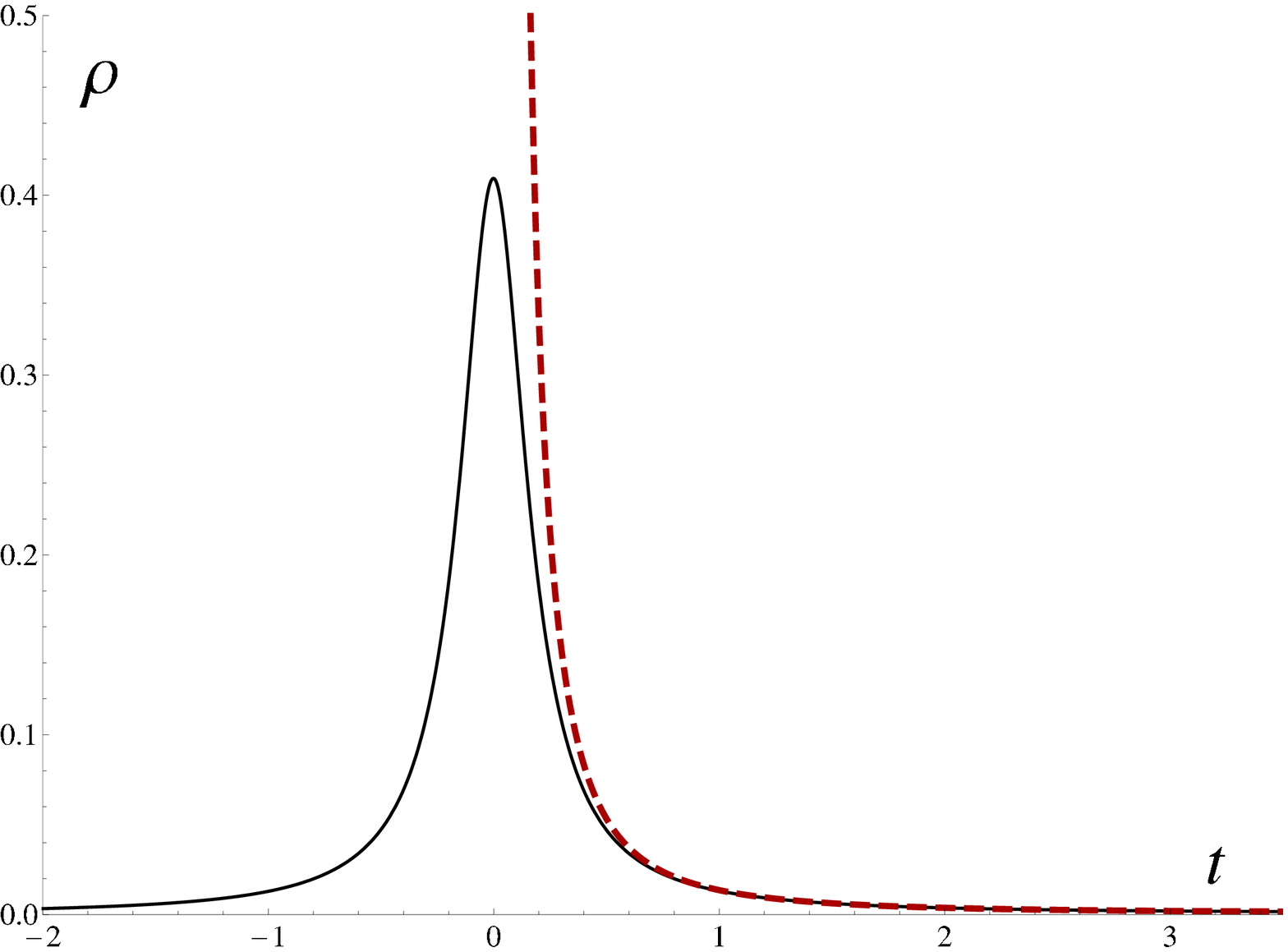}
\caption{Evolution of the scale factor and energy density is shown in the Planck regime for LQC (solid curve) and the classical theory (dashed curve), for a single scalar field with potential (\ref{pot1}).  Values of the parameters of the potential are $V_0 = 0.001$ and $k=0.5$, and the units are chosen as Planck units. The scale factor in the classical theory vanishes, causing energy density to diverge. In LQC, energy density reaches a maximum value and the scale factor bounces.} %The plot on the left hand side shows the evolution of the scale factors in the bounce regime. Long term variation of the scale factors is shown on the right hand side.  The late-time linear relationship of $\ln a$ versus $ \ln t$ indicates the familiar power law as an asymptotic solution. }
\end{figure}

\begin{figure}[tbh!]
%\begin{center}
%\label{fig1}
\includegraphics[angle=0,width=0.445\textwidth]{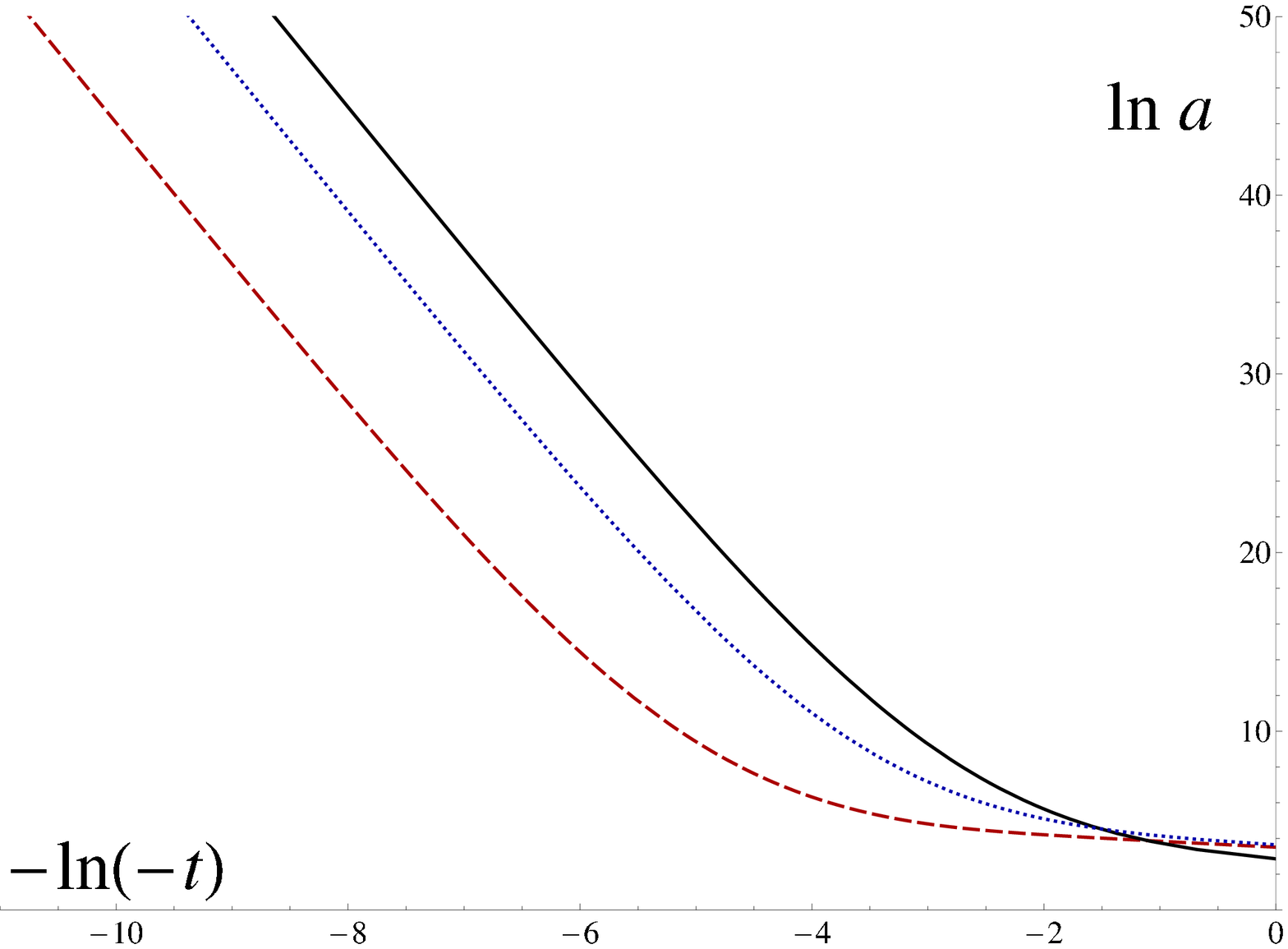}
\includegraphics[angle=0,width=0.455\textwidth]{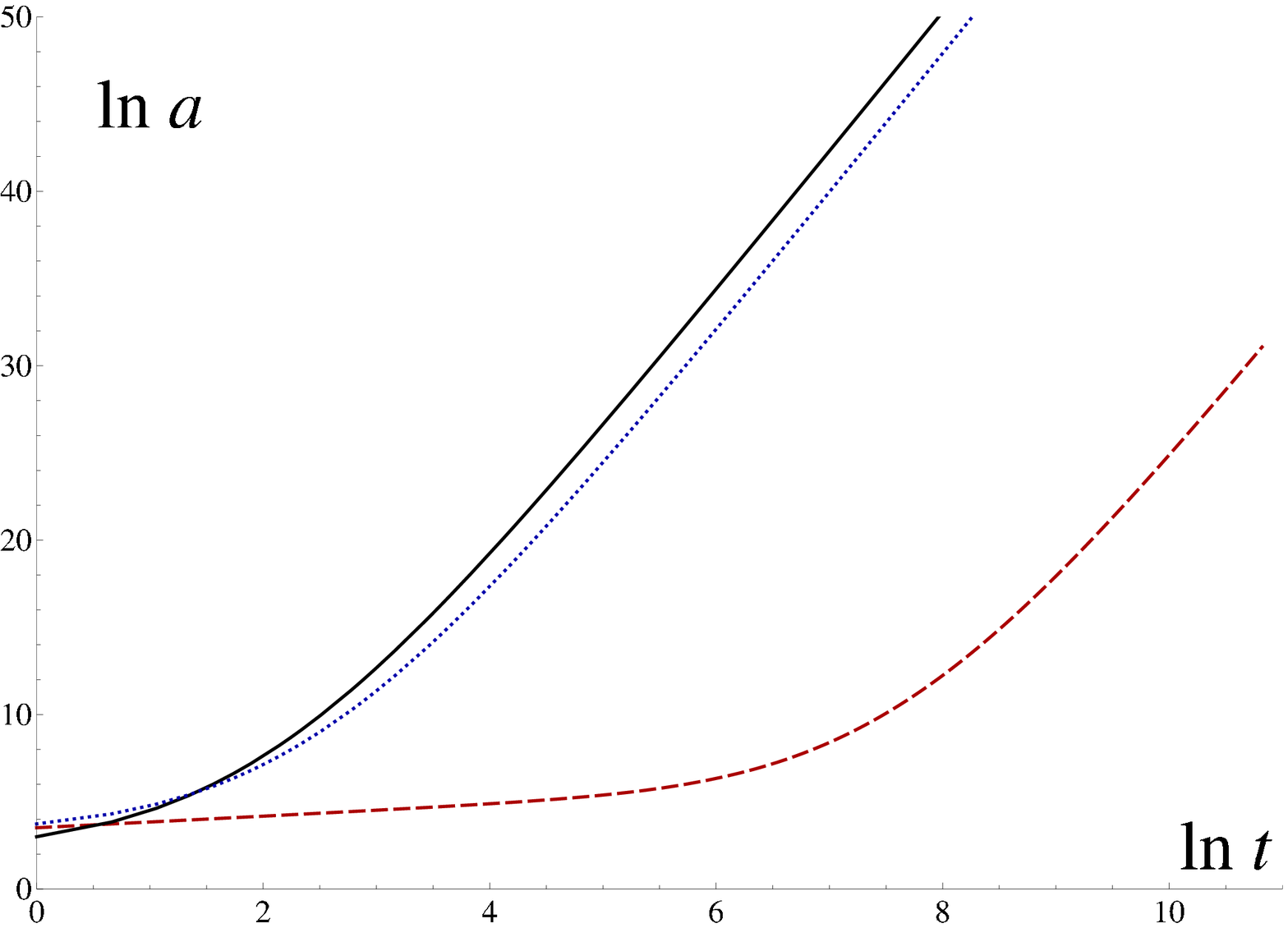}
\caption{Long term evolution of the scale factor is shown for different initial conditions in the pre-bounce (left plot) and post-bounce (right plot). For the chosen initial conditions, bounce occurs at $t=0$ (not shown due to logarithmic scale). At very early (pre-bounce) and late (post-bounce) times, the classical power-law evolution is obtained.}
%Evolution of the scale factor, in Planck regime, in LQC (solid curve) and the classical theory (dashed curve) is shown for a single scalar field with potential (\ref{pot1}).  Values of the parameters of the potential are $V_0 = 0.001$ and $k=0.5$, and the units are chosen as Planck units. The scale factor in the classical theory vanishes in the classical theory causing energy density to diverge. In LQC, energy density reaches a maximum value and the scale factor bounces.} %The plot on the left hand side shows the evolution of the scale factors in the bounce regime. Long term variation of the scale factors is shown on the right hand side.  The late-time linear relationship of $\ln a$ versus $ \ln t$ indicates the familiar power law as an asymptotic solution. }
\end{figure}

In the classical theory, dynamical solutions for the potential (\ref{pot1}) are generically singular. In contrast, in LQC, all the solutions with the above potential turn out to be non-singular. The differences between the classical theory and LQC only become significant in the Planck regime. In the regime $\rho \ll \rcr$, the modified Friedmann and Raychaudhuri equations (\ref{modfried},\ref{modrai}) approximate their classical counterparts (\ref{fried},\ref{rai}), and, as expected, solutions in classical theory and LQC agree. This agreement occurs at late time after the bounce, as well as early times before the bounce, where LQC solutions furnish the same scaling property as in General Relativity (GR). A comparison of the solutions in the classical theory and LQC is illustrated in Fig. 1, and the early and late time behavior of solutions is shown in Fig. 2. In Fig. 1, we see that in the classical theory, the scale factor vanishes in a finite time in the past evolution of an expanding branch. This causes the energy density and the curvature invariants to diverge, and thus the physical evolution breaks down. In contrast, in LQC, starting with the initial conditions corresponding to the classical solutions at late times, we find that the dynamical solutions approximate the classical ones for a long time, however there are significant departures from the classical trajectory when energy density of the scalar field approaches $\rmax$. Using eqs.(\ref{modfried}) and (\ref{modrai}), we find that the scale factor  in LQC bounces when $\rho = \rmax$, and the backward evolution continues beyond the classical singularity. The linear relationship of $\ln a$ versus $ \ln t$ seen in Fig. 2 indicates the familiar power law behavior of the classical theory. Quantum gravitational effects encoded in eqs. (\ref{modfried}) and (\ref{modrai}), thus lead to a past complete non-singular power-law inflationary model. These turn out to be robust features of the extensive numerical simulations performed with potential (\ref{pot1}). In Fig. 3, we show the plots for the relative errors in the effective Hamiltonian constraint (\ref{effham}) for the solution discussed in Fig. 1. We see that the error in the effective Hamiltonian constraint is very small compared to the individual terms, $H_{\rm{ grav}}$ and $H_{\rm{matt}}$. The behavior of relative errors shows that the effective Hamiltonian constraint is satisfied almost perfectly. (Relative errors remained similar in all the numerical solutions discussed in our analysis).

\begin{figure}[tbh!]
%\begin{center}
\label{phaseplot}
\includegraphics[angle=0,width=0.45\textwidth]{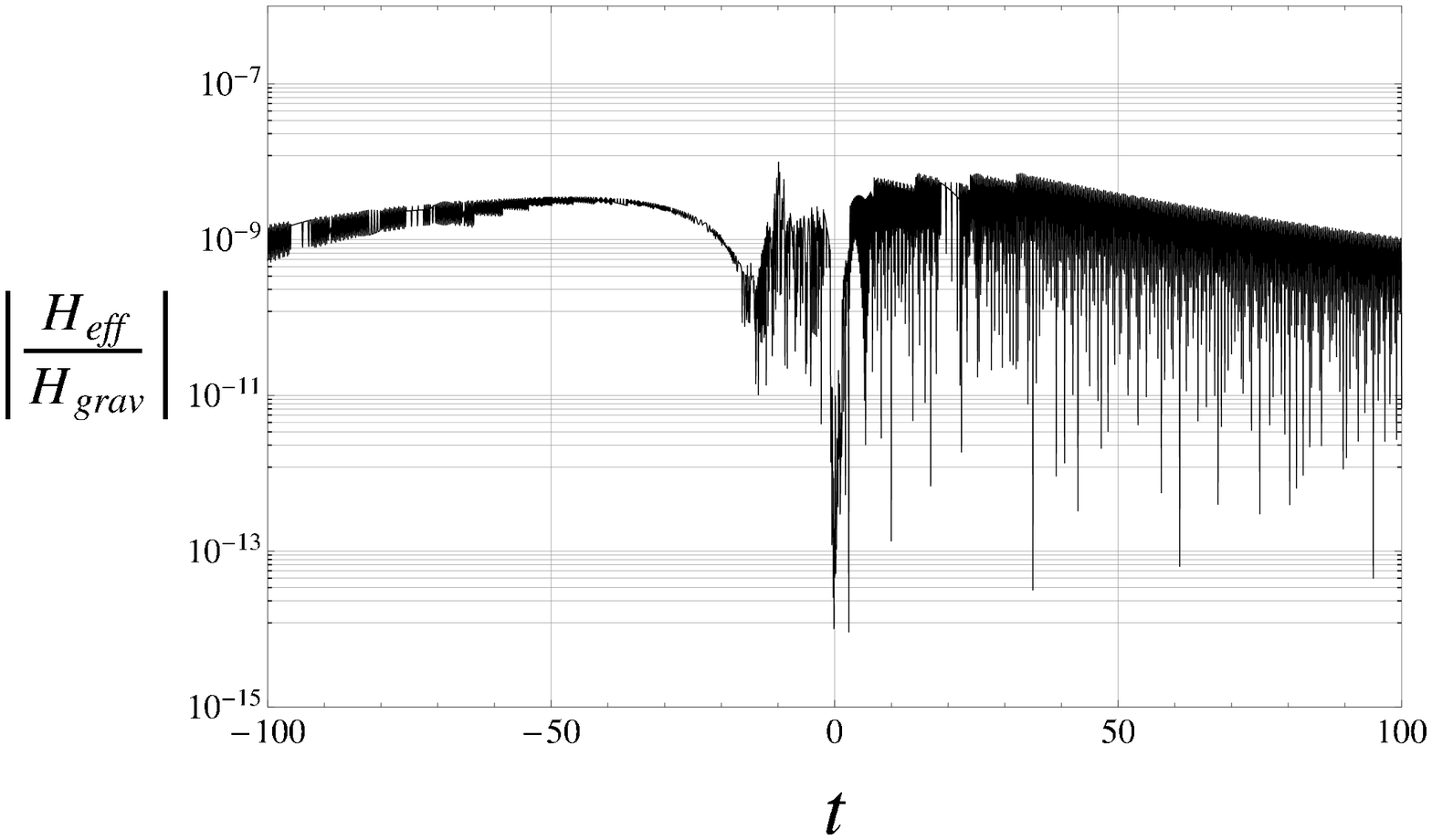}
\includegraphics[angle=0,width=0.45\textwidth]{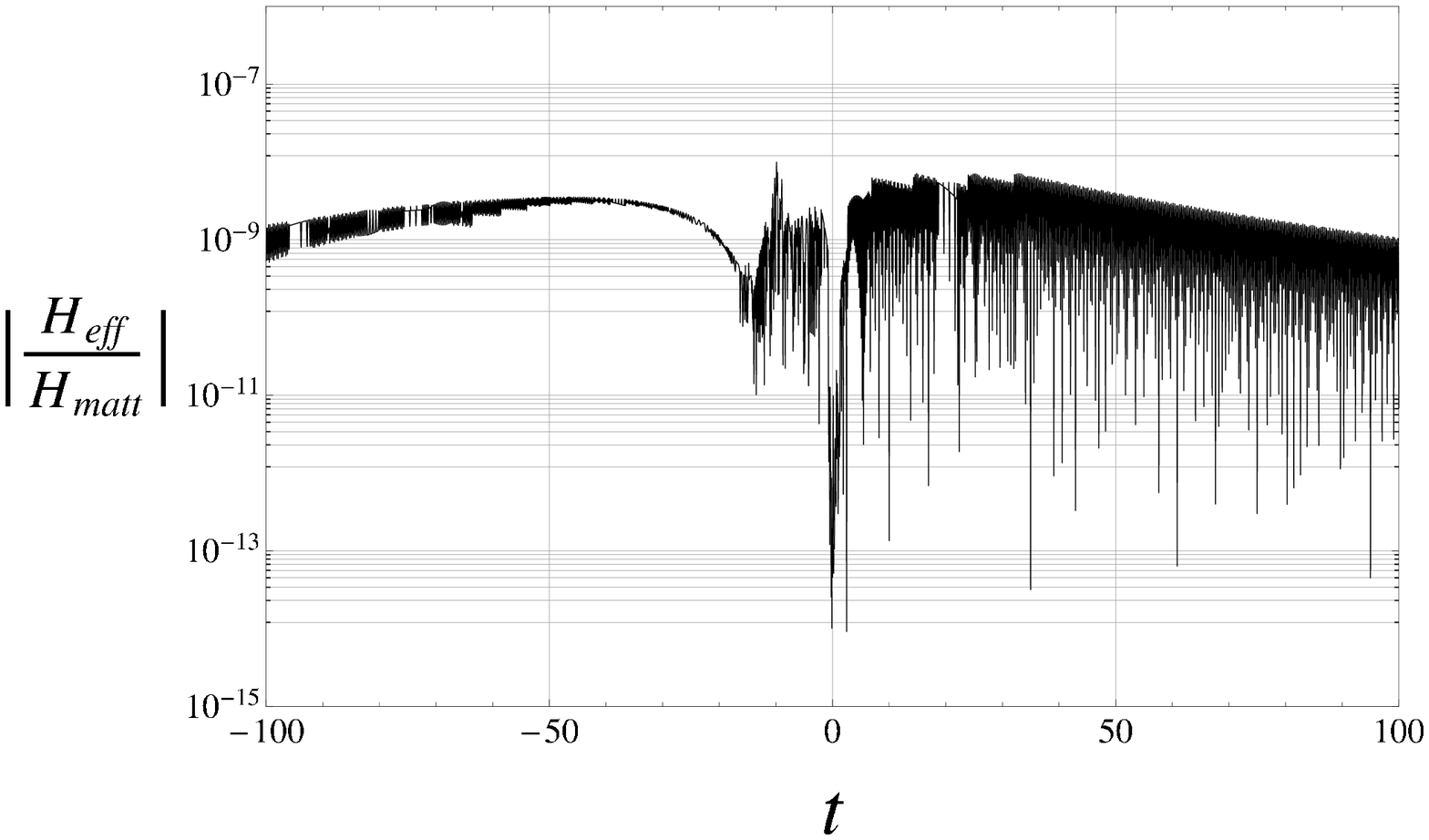}
\caption{An analysis of the error for the solution with the solid curve in Fig. 1 is shown. The ratios $|H_{\mathrm{eff}} / H_{\mathrm{grav}}|$ and $|H_{\mathrm{eff}}/H_{\mathrm{matt}}|$ remain small during the evolution.} %
\end{figure}

Let us now examine the behavior of matter variables, $\phi$ and $\dot \phi$, in the dynamical evolution. This can be understood by a phase space plot such as in Fig. 4, where we discuss the trajectories for the same values of parameters as in Fig. 1. Depending on the sign of the initial velocity $\dot \phi$, classical curves approach the classical attractor in the expanding branch at large $\phi$ at late times from the upper and the bottom halves of $\phi - \dot\phi$ plane. At early times, all of the classical dynamical trajectories, approach the regime where $|\dot \phi|$ diverges, and near the singularity the phase space trajectories asymptotically behave as for the massless scalar model. Similar behavior of the trajectories exist for the contracting branch, where in the forward evolution classical curves starting from a repeller at very early times, approach the massless scalar regime near the big crunch singularity. In LQC, the behavior of trajectories in $\phi - \dot \phi$ plane has some important differences with those in the classical theory. First, unlike in the classical theory, trajectories in LQC are bounded in $|\dot \phi|$. This bound is a direct consequence of the upper bound on the allowed energy density $\rcr$ in LQC. Second, in LQC, dynamical trajectories starting from the pre-bounce repeller in contracting branch approach the late time post-bounce attractor in the expanding branch. Using extensive numerical simulations, we found that dynamical trajectories in LQC first converge towards a non-classical curve in the post-bounce phase in the regime  when energy density is $\rho \lesssim \rmax/2$, before they approach the classical attractor.   This can be seen from the behavior of LQC trajectories in the upper half plane in Fig. 5, where small differences in the curve to which LQC trajectories initially converge and classical theory are visible when $\rho \sim \rmax/2$. The difference becomes negligible when the energy density decreases. A similar behavior exists for the repeller in the forward evolution of the contracting branch (in the lower half-plane of $\phi-\dot \phi$ plot).
%At early times (before bounce), and late times (after bounce), these attractors approach the backward attractor in the classical contracting branch and the forward attractor in the classical expanding branch.
%Another notable feature of the phase space trajectories apparent from the phase space plot (Fig. 4), is the existence of an attractor, both in the post-bounce and pre-bounce regime, different from the ones in the classical theory, to which LQC trajectories approach.
%Thus, quantum gravitational effects in LQC, bridge the two disjoint attractors in the classical theory.

\begin{figure}[tbh!]
%\begin{center}
%\label{phaseplot}
\includegraphics[angle=0,width=0.5\textwidth]{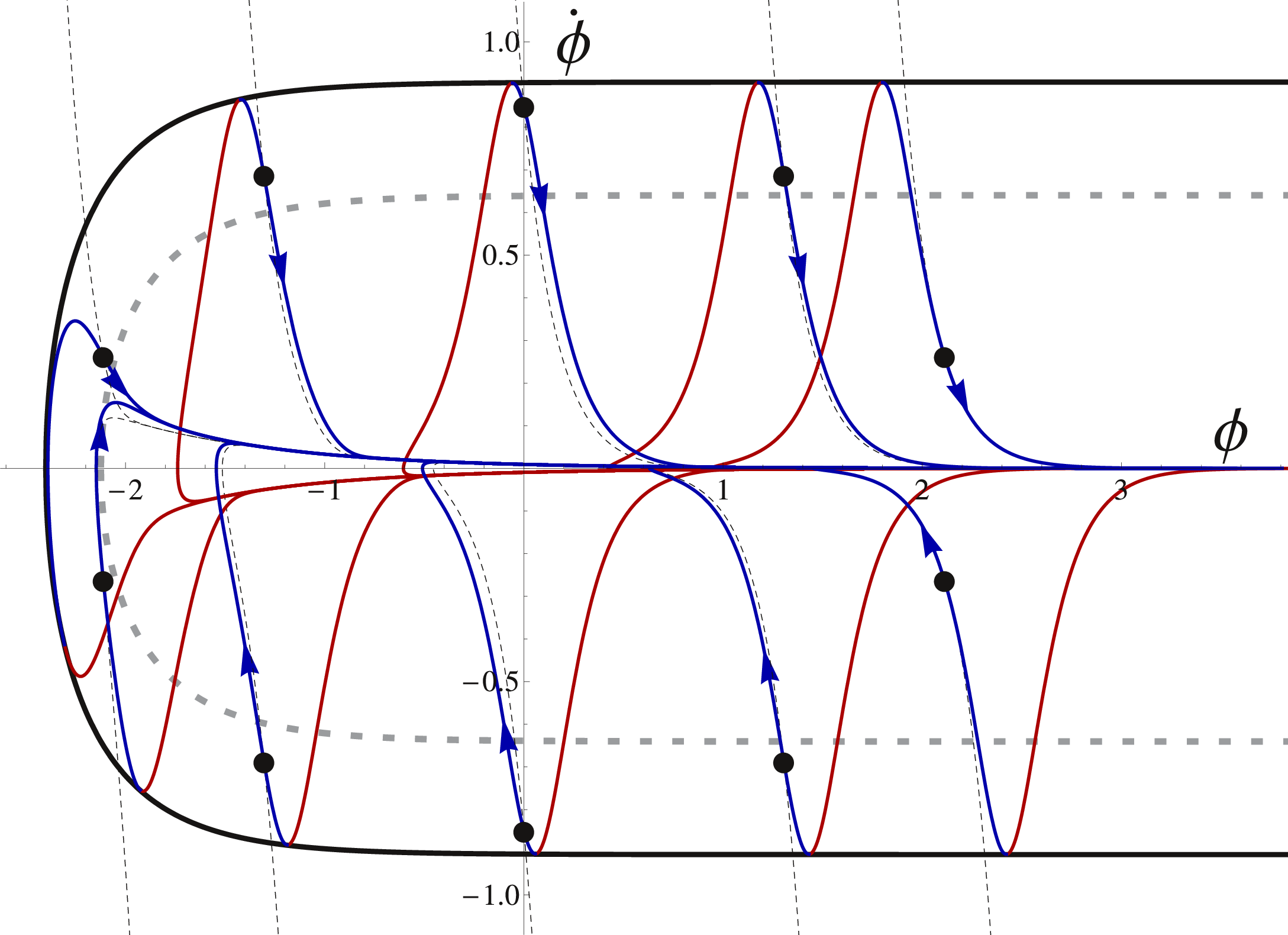}
\caption{Phase plots with several initial conditions chosen  in the $\phi-\dot{\phi}$ plane for a field with exponential function given by $V_0 = 0.001$, $k=0.5$. Initial conditions appear as solid dots on the solution curves, and arrows denote the direction of forward time evolution. The thick solid contour represents a level curve of energy density $\rho = \rcr$. The dashed contour represents energy density $\rho = \rcr/2$.}
\end{figure}

\begin{figure}[tbh!]
%\begin{center}
%\label{phaseplot}
\includegraphics[angle=0,width=0.5\textwidth]{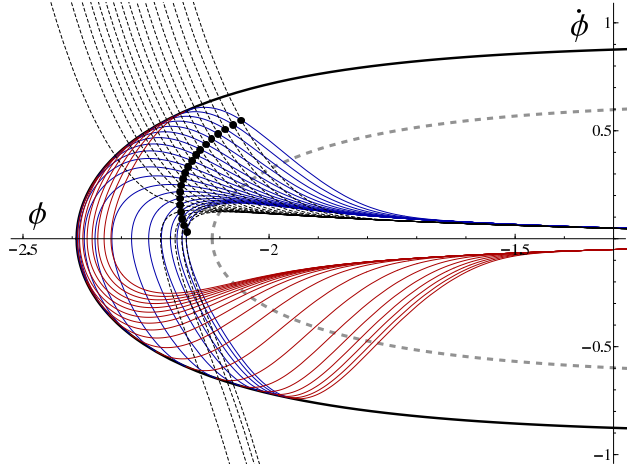}
\caption{This plot shows a zoom of the phase space trajectories in the Planck region. The field and graphical notation is the same as in Fig. 3. Initial conditions correspond to densities between $0.6 \rcr$ and $0.8 \rcr$. For the energy densities close to $\rmax/2$, LQC trajectories approach an attractor in the forward evolution, which approximates the classical attractor when energy density becomes much smaller than $\rmax/2$.} %A similar behavior exists for the backward evolution in the lower half plane. }
\end{figure}

 We now consider some more detailed properties of the phase space trajectories in the Planck regime. In Fig. 5,  initial conditions are provided, all corresponding to positive $\dot \phi$, in a narrow range of energy density in the super-inflation regime in LQC $(\rcr > \rho > \rcr/2$). We find that in the backward evolution, depending on the initial conditions on $\phi$ and $\dot \phi$, trajectories in LQC encounter a quantum bounce at $\rho = \rcr$, where $\dot \phi$ can be positive or negative. The magnitude of $\dot \phi$ can take any value, including zero, at the bounce of the scale factor below the maximum bound set by $\rmax$. Thus, the quantum bounce can be kinetic (i.e. $w = P/\rho \sim 1$) or potential dominated (i.e. $w \sim -1$). In the numerical simulations we found that if  initial conditions are specified away from the bounce, trajectories which lead to kinetic dominated bounce are easier to find than those which lead to the potential dominated bounce. Thus in Fig. 5, trajectories approaching $\dot \phi \sim 0$ at the bounce, in the backward evolution from the initial conditions specified in Planck regime, are fewer than those which correspond to $\dot \phi \nsim 0$. We emphasize that this behavior strongly depends on the specification of the initial conditions in the phase space plot. Finally, we note that since in LQC $\phi$ and $\dot \phi$ can take on a range of values at the bounce, the equation of state for the scalar field at the bounce can take on values between $w=1$ and $w=-1$.

\begin{figure}[tbh!]
%\begin{center}
%\label{phaseplot}
\includegraphics[angle=0,width=0.5\textwidth]{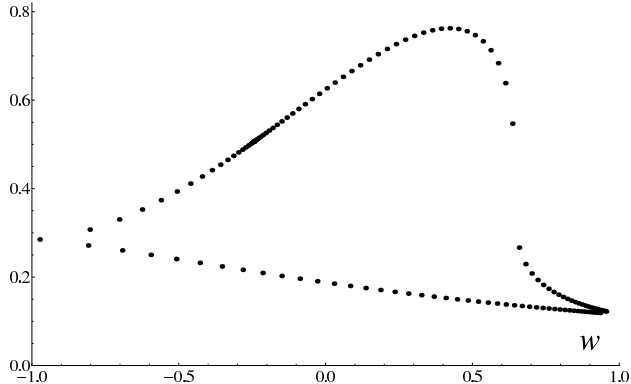}
\caption{The amount of super-inflation (in e-foldings) vs $w$, the equation of state, at the time of bounce for a field with steepness parameter $k=1$. A wide range of $w$ values is obtained by choosing many initial conditions along the curve defined by $\rho=0.95\rcr$. }%Values of $V_o$ not imp.}
\end{figure}

Having noted the main features of the dynamics and the phase space trajectories, we now turn our attention to understand the super-inflation phase in this model. This phase occurs when energy density of the scalar field is $\rmax > \rho > \rmax/2$, where $\dot H > 0$, and contributes to additional e-foldings originating purely due to quantum gravity effects. A pertinent question is whether these e-foldings are large and under what conditions.  It turns out that the number of e-foldings depend  on the equation of state at the bounce, as well as the parameter $k$ in a non-trivial way.\footnote{These do not depend on the value of $V_o$, a change in which only causes a trivial shift in $\phi$.} For each given value of $w$ with a particular value of $k$,  there exist two values of the number of e-foldings in the super-inflationary regime, as is depicted in Fig. 6. In the special cases of $w=-1$ at the bounce, these two values coincide. This behavior can be understood by noting that for the trajectories in the $\phi-\dot \phi$ plane, for a fixed value of $\phi$ at the bounce, energy density can be equal to $\rmax$ for two values of $\dot \phi$ with same magnitude but different signs. For different signs of $\dot \phi$ at the bounce, the evolution of energy density and Hubble rates in the regime $\rmax > \rho > \rmax/2$ is different. This difference in evolution causes, two values of the number of e-foldings for the same equation of state, one corresponding to positive  $\dot \phi$ at the bounce, and the other to the negative $\dot \phi$. For a given value of $w$ at the bounce, the magnitude of the difference depends on the value of the parameter $k$.

\begin{figure}[tbh!]
\label{fourfig1}
\includegraphics[angle=0,width=0.8\textwidth]{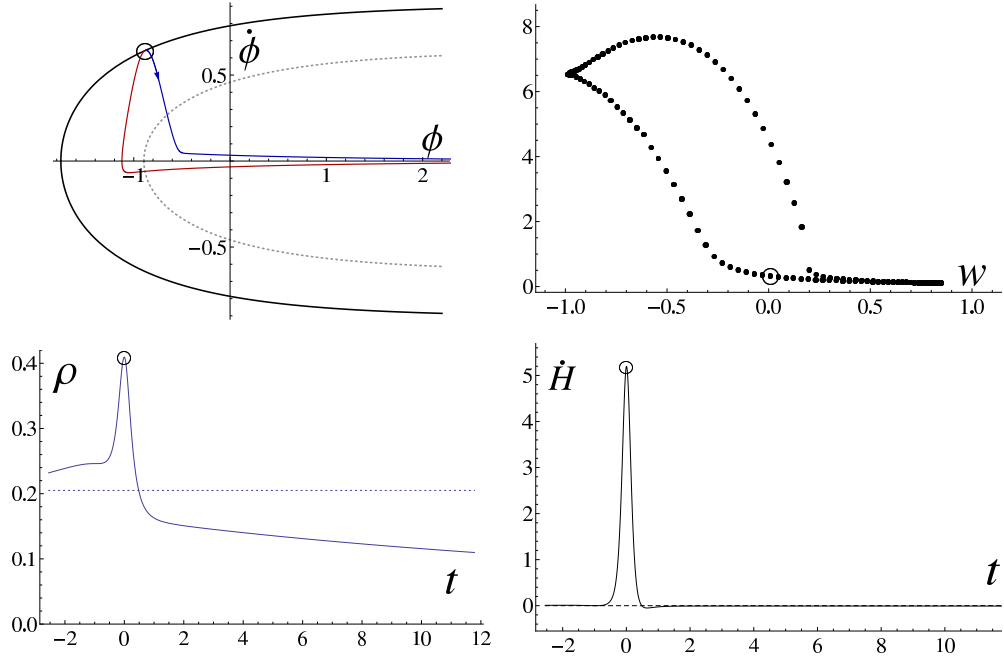}
\caption{Phase space trajectory and the behavior of energy density and time derivative of Hubble rate, along with the number of e-foldings versus equation of state are shown for $k = 0.16$ and $V_o = 0.1$. A circle on the curve denotes the value at the bounce. Each dot in the plot of the number of super-inflationary e-foldings corresponds to an initial condition taken at $\rho = 0.999 \rmax$.  }
\end{figure}

\begin{figure}[tbh!]
\label{fourfig2}
\includegraphics[angle=0,width=0.8\textwidth]{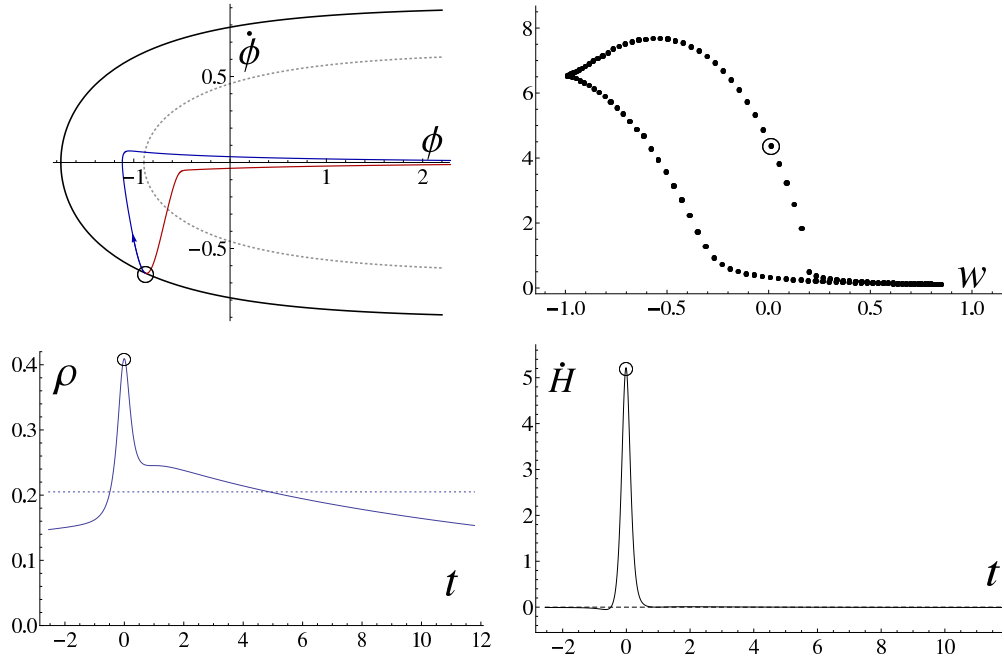}
\caption{For the same values of $V_o$ and $k$  as in Fig. 7, the phase space trajectory, behavior of energy density and $\dot H$, and the value of equation of state at the bounce are shown for different initial conditions. Since $\dot \phi < 0$ at the bounce, the number of e-foldings during super-inflation are higher than in Fig. 7. }
\end{figure}

This behavior is illustrated by the Figs. 7 \& 8, where for a fixed value of the parameters in the potential, we have compared the phase space trajectory, the behavior of energy density and Hubble rate, and the amount of super-inflation versus the equation of state at bounce. This is done for two sets of initial conditions leading to the bounce at the same value of $\phi$, and $\dot \phi$ with equal magnitude but opposite signs. We find that in the case for $\dot \phi > 0$ at the bounce (Fig. 7), the number of e-foldings in the super-inflation regime is less than 0.5, whereas  for the case when $\dot \phi < 0$ (Fig. 8), it is approximately equal to 4. From the Fig. 7 \& 8, we see that this difference is due to the relatively longer duration in which $\rho > \rmax/2$ when $\dot \phi$ is negative at the bounce, compared to when it is positive. The cause of the difference in this duration is ultimately tied to the fact that, in the forward evolution, trajectories for which the bounce occurs in the lower half of the $\phi-\dot \phi$ plane cover a longer path in the phase plane to reach $\rmax/2$ and merge with the forward attractor (which lies in the upper half plane) than those which bounce in the upper half plane. %{\Large{Why the forward attractor always lies in the upper half plane?}}
In Fig. 9, we illustrate the special case of $w=-1$ at the bounce. In this case, the number of e-foldings during super-inflation turn out to be more than those for the initial conditions in Figs. 7 \& 8, where the equation of state at the bounce is approximately zero. A comparison of the plots of $\rho$ and $\dot H$ in Figs. 7, 8 \& 9, reveals that for the case of bounce with $w = -1$, the variation of energy density and rate of change of the Hubble rate favors a longer duration of super-inflation and hence a larger number of e-foldings.

Interestingly, as the value of the parameter $k$ is changed, the maximum number of e-foldings in the super-inflationary phase changes. Further, the value of the equation of state at the bounce which yields the maximum number of super-inflationary e-foldings also depends on $k$, though as expected it is independent of $V_o$. From Fig. 6, we find that for the parameter $k$ equal to unity,  there are approximately 0.8 e-foldings as the maximum number of e-foldings due to super-inflation when the equation of state at the bounce is $w \sim 0.5$. Whereas, for $k=0.16$ the maximum number of e-foldings are approximately 8 when $w \sim -0.5$ at bounce. We performed a detailed analysis of this behavior for various sets of initial conditions with different values of $k$, and the results are represented in Fig. 10. Two main features appear from this analysis:\\

\noindent
(i) The dependence of the maximum number of e-foldings in super-inflationary phase is non-linearly related to the value of $k$.  As $k$ decreases, i.e. the potential (\ref{pot1}) becomes shallower, the maximum number of e-foldings during super-inflationary phase increases. In our simulations, we found that for $k = 0.06$, the number of e-foldings during super-inflation can be as high as 50. As the potential becomes steeper, i.e. $k$ increases, the maximum number of e-foldings during super-inflation decreases.\\

\noindent
(ii) The equation of state at the bounce yielding the maximum number of e-foldings during super-inflation depends on $k$ in a non-linear way. For steep potentials, maximum number of e-foldings in super-inflation occur when the bounce is not potential dominated. As this parameter decreases, the potential becomes shallower, and the maximum number of e-foldings during super-inflation occur for equation of the state at the bounce approaching $-1$. Fig. 10 suggests that the maximum number of e-foldings occur for $w=-1$ as $k$ approaches zero.

\begin{figure}
\label{fourfig3}
\includegraphics[angle=0,width=0.8\textwidth]{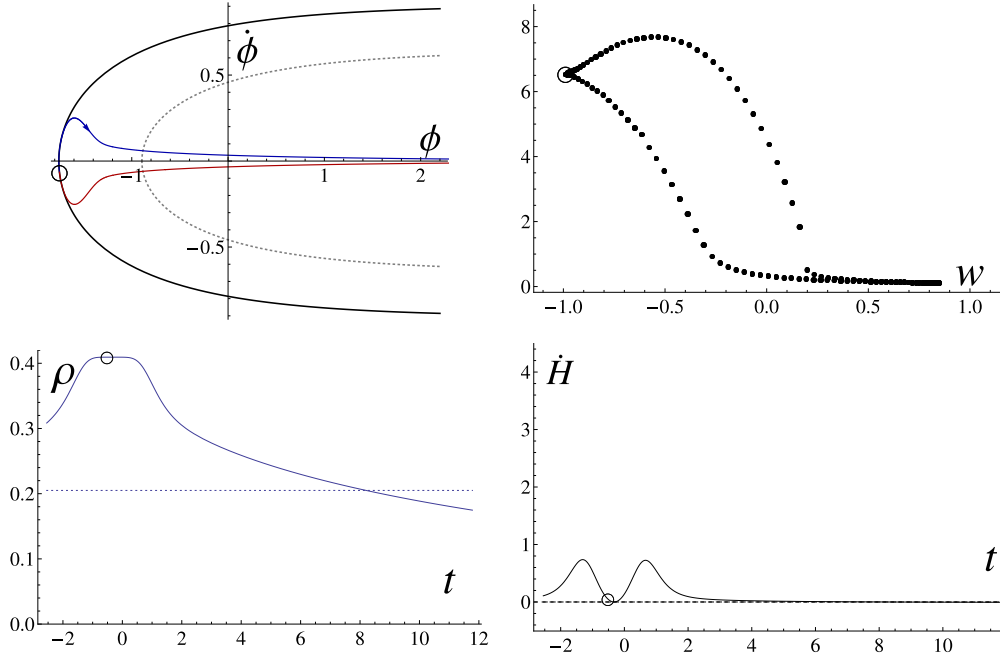}
\caption{The plot shows the way variation in energy density and $\dot H$, due to the behavior of $\dot \phi$ affects the number of e-foldings during super-inflation. The parameters of potential are same as in Fig. 7, and only initial conditions are changed.  The bounce occurs at $w = -1$, which yields more (but not the maximum) e-foldings in comparison to initial conditions in Fig. 7 \& 8.}
\end{figure}

%\begin{figure}[tbh!]
%\begin{center}
%\label{phaseplot}
%\includegraphics[angle=0,width=0.5\textwidth]{figure10.eps}
%\caption{The maximum number of e-folding during super inflation is shown versus equation of state  at the time of bounce.  for a field with %$k=0.2$. A wide range of $w$ values is obtained by choosing many initial conditions along the curve defined by $\rho=0.95\rcr$}
%\end{figure}

\begin{figure}[tbh!]
%\begin{center}
%\label{phaseplot}
\includegraphics[angle=0,width=0.45\textwidth]{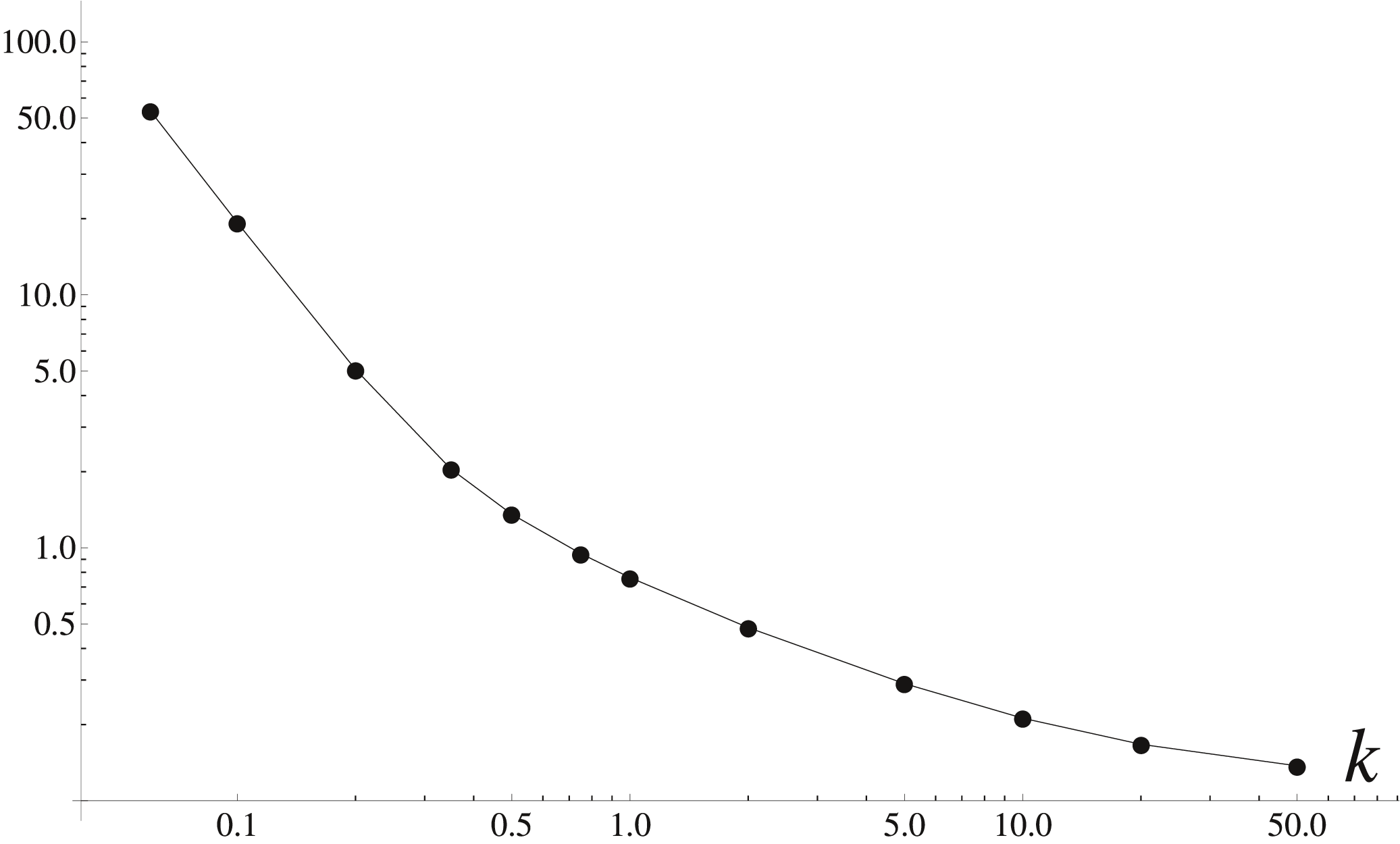}
\includegraphics[angle=0,width=0.45\textwidth]{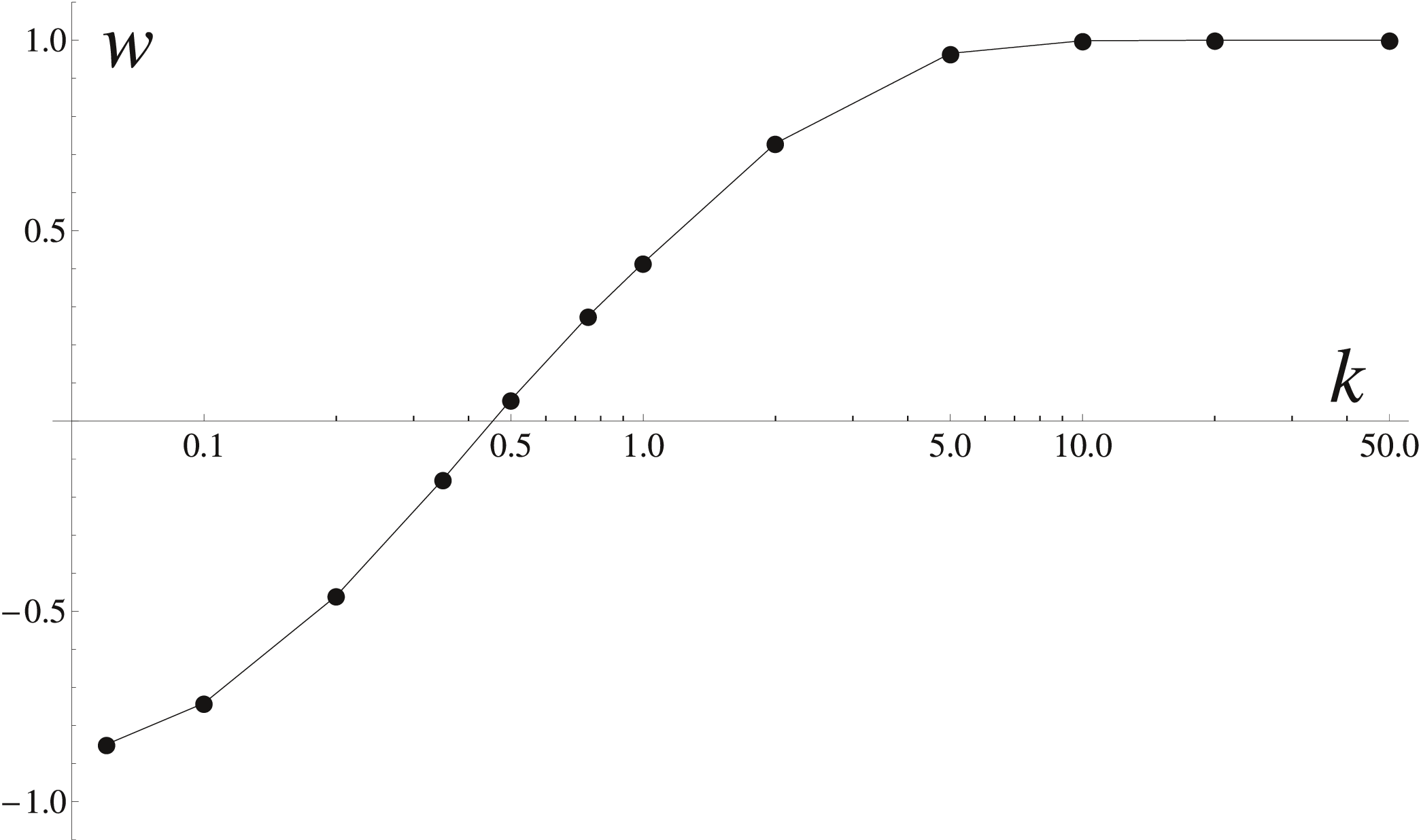}
\caption{Left plot shows the dependence of the maximum number of e-folding during super-inflation (plotted along y-axis). This number increases as $k$ decreases. The right plot shows the relationship between the equation of state at the bounce which produces the maximum number of e-foldings during super-inflation and the parameter $k$.}
\end{figure}

%\begin{figure}[tbh!]
%\begin{center}
%\label{phaseplot}

%\caption{The maximum number of e-folding during super inflation is shown versus equation of state  at the time of bounce.  for a field with %$k=0.2$. A wide range of $w$ values is obtained by choosing many initial conditions along the curve defined by $\rho=0.95\rcr$}
%\end{figure}

%\begin{figure}[tbh!]
%\begin{center}
%\label{phaseplot}
%\includegraphics[angle=0,width=1.0\textwidth]{figure13.eps}
%\includegraphics[angle=0,width=0.5\textwidth]{figure14.eps}
%\caption{The maximum number of e-folding during super inflation is shown versus equation of state  at the time of bounce.  for a field with %$k=0.2$. A wide range of $w$ values is obtained by choosing many initial conditions along the curve defined by $\rho=0.95\rcr$}
%\end{figure}

\section{Assisted inflation}
As discussed earlier, for the potential of the form (\ref{pot1}), if $k^2 > 2$, a single scalar field does not lead to inflationary dynamics in GR. However,  multiple fields $\phi_i$, each with an exponential potential (\ref{pot1}) with a steepness parameter $k_i$, which couple through Hubble expansion, can lead to inflation even if $k_i^2 > 2$ for each field. This novel mechanism, in which individual fields cooperate to produce inflation when none of them could have done so individually, forms the basis of assisted inflation \cite{assisted1}. The combined potential of multiple fields, $V$, is given by
%We label the potential of each of the field $\phi_i$ as $V_i(\phi_i)$ (with same $V_o$), which leads to the combined potential of multiple fields as
\be
V = \sum_i \,V_i(\phi_i) = \sum_i\, V_{o} \, e^{-\sqrt{8 \pi G} \,k_i \phi_i} ~.
\ee
where $V_i$ denotes the potential of each of the field $\phi_i$, and $V_o$ is considered to be the same for all fields. In such a scenario, the dynamical equations for the gravitationally coupled multiple fields can be re-expressed as those of an effective single field $\phi$ with an equivalent $k_{\rm eq}$ given by \cite{assisted1}
\be
\f{1}{k_{\rm eq}^2} = \sum_i \f{1}{k_i^2} ~.
\ee
Thus, even though each $k_i^2 > 2$, $k_{\mathrm{eq}}^2$ can be less than 2. If so, the multiple field configuration can successfully lead to inflation with a power law expansion $a \propto t^{n_{\mathrm{eq}}}$ with
$n_{\mathrm{eq}} = $ $\displaystyle\sum\limits_i$ $2/k_i^2 > 1$.  In this way different fields assist each other in causing inflation \cite{assisted1}. In fact, as the number of fields increases, inflation becomes possible for larger values of the individual $k_i$'s, and the effective potential $V$ becomes shallower. Since the multi-field system in assisted inflation reduces to an effective single field scenario, various results of the previous section are applicable here. In the classical dynamics, in the forward evolution the fields evolve with energy densities in fixed ratios, leading to a scaling solution which is an attractor \cite{assisted2}. In the backward evolution, as the past singularity is approached, dynamical trajectories in $\phi - \dot \phi$ plane approach those of the massless scalar fields and the potential $V$ becomes insignificant near the singularity. On the other hand, in LQC, singularity is avoided irrespective of the choice of the number
of fields, parameters $V_o$ and $k_i$'s and the initial conditions. However, the behavior of each field near the bounce is now determined collectively by the steepness parameters $k_i$ and the initial conditions of all the fields.
%\be
%n = \sum_i \,\f{2}{k_i^2} = \sum_i n_i
%\ee

\begin{figure}[tbh!]
%\begin{center}
\label{multifield1}
\includegraphics[angle=0,width=0.5\textwidth]{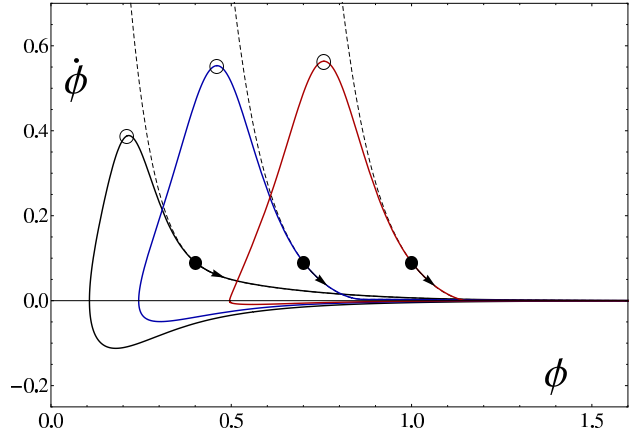}
\caption{The multi-field dynamics of 3 fields $\phi_1$, $\phi_2$ and $\phi_3$ is shown (ordered left to right). The fields have potential parameters as: $V_{o} = 0.1, k_1 = 1.5$, $k_2 = 1.75$, and $k_3 = 2$ respectively. The initial conditions are $\phi_1(0) = 0.4, \phi_2(0) = 0.7, \phi_3(0) = 1$ and $\dot \phi_1(0) = \dot \phi_2(0) = \dot \phi_3(0) = 0.091$ in Planck units, and their location is denoted by thick black dots on the curves. The arrows indicate the forward time direction, and the dashed lines represent classical solutions to the same initial conditions. The bounce occurs at the three points marked by an open circle. }
\end{figure}
\begin{figure}[tbh!]
%\begin{center}
\label{mf2}
\includegraphics[angle=0,width=0.5\textwidth]{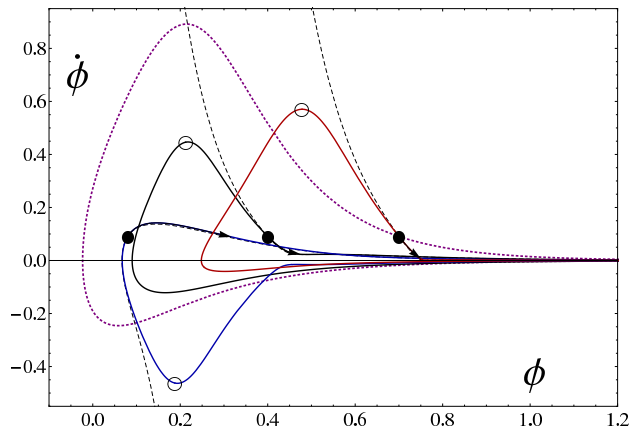}
\caption{Plot of the dynamical trajectories in the classical theory (dashed curves) and LQC (solid curves) are shown for the three field system with choice of parameters as in Fig. 11. The initial conditions for $\phi_1, \phi_2$ and $\phi_3$ are similar to those in Fig. 11, with the only different values being $\phi_2(0) = 0.08$ and $\phi_3(0) = 0.7$. (in Planck units). The trajectories for fields are plotted from left to right in the order: $\phi_2$, $\phi_1$, $\phi_3$. The thick dashed curve shows the dynamical trajectory in LQC for the field $\phi_3$ treated as in single field evolution.}
\end{figure}

To understand the dynamics in a multi-field scenario, we performed various numerical simulations with a system of three fields. An example of the dynamical trajectories in the $\phi-\dot \phi$ plane are shown in Fig. 11 where we consider steepness parameters as $k_1 = 1.5, k_2 = 1.75$ and $k_3 = 2.0$. The equivalent steepness parameter for the single field case turns out to be $k_{\rm{eq}} \approx 0.99$. Thus, even though each of the three scalar fields can not individually lead to inflation, the effective steepness parameter is such that  assisted inflation is possible. From Fig. 11 we see that in the backward evolution of the expanding branch, dynamical trajectories in LQC do not follow classical trajectories in to the singularity but bounce and turn-around to approach the attractor in the backward evolution of the contracting branch. In the forward evolution, dynamical trajectories in LQC approximate the classical trajectories when $\rho \ll \rmax$ and approach the late-time attractor of the multi-field system in the classical theory \cite{assisted2}. Unlike in LQC, in the classical theory, the attractor in the forward evolution of the expanding branch is disjoint from the repeller in the forward evolution of the contracting branch due to the presence of the big bang singularity.  We find that the field $\phi_1$ approaches $\phi = 0$ in the forward evolution more slowly than $\phi_2$ and $\phi_3$. In terms of the single field evolution, this behavior can be understood by noting that the %initial potential energy of $\phi_1$ is greater than that of $\phi_2$ and $\phi_3$, and further
the effective steepness parameter $k_{\rm eq}$ of the system is closer to $k_1$ than  $k_2$ or $k_3$.

The effect of the choice of initial conditions in multi-field scenario can be seen by comparing Fig. 12 with Fig. 11. In Fig. 12, we consider the same set of parameters and initial conditions as in Fig. 11 but with a change in the initial potential energy of the fields $\phi_2$ and $\phi_3$. We find that the new initial conditions cause a non-trivial change in the way different fields approach the scaling solution at late times (after the bounce), and also at early times in the pre-bounce phase. Unlike the case in Fig. 11, the bounce of the scale factor occurs when $\dot \phi_2$ is negative, and the kinetic energy of the field is less dominant than before, though the implications these changes have on the system is much more complicated than in the single field cases explored earlier. Fig. 12 also contrasts the behavior of the single field case with that of multi-fields. For the field $\phi_3$, we show the dynamical trajectory (thick dashed curve) for the same initial conditions and parameters but evolved as an individual field. It is clearly seen that the attractor and the location of the bounce differ between the single field and multi-field cases with the same initial conditions, due to the additional contribution to Hubble damping by fields $\phi_1$ and $\phi_2$ in the latter case.

%\begin{figure}[tbh!]
%\begin{center}
%\label{mf2}
%\includegraphics[angle=0,width=0.5\textwidth]{figure16.eps}
%\includegraphics[angle=0,width=0.45\textwidth]{figure18.eps}
%\caption{These two figures show the evolution of the individual field densities by looking at the ratio of each to the total density of the system. Time is shown as ln(t), and so the solutions have been adjusted so that t=1 corresponds to the specification of initial conditions, which appears at ln(t)=0 on the graph. Initial values and fields are taken from the corresponding figures, figures 7 and 8. Fields can easily be identified using the fact that on the right side of each graph where the asymptotic solution is reached, the field densities are $\rho_a, \rho_b,\rho_c$ in Decreasing order (A on top, etc.)}
%\end{figure}

\begin{figure}[tbh!]
%\begin{center}
%\label{multifield2}
\includegraphics[angle=0,width=0.5\textwidth]{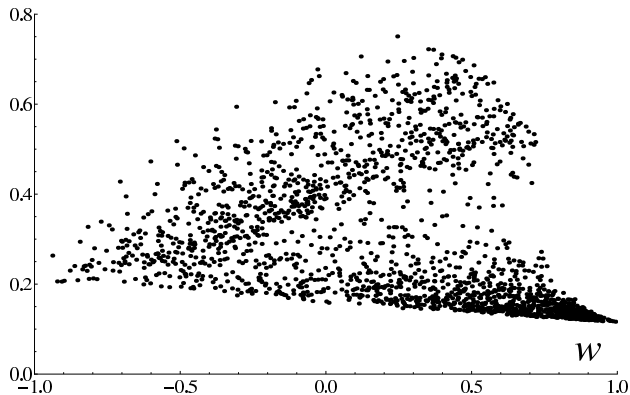}
\caption{The number of e-foldings obtained during super-inflation versus equation of state for a three field system in assisted inflation is shown. Parameters are $k_1=1.5$, $k_2=1.6$, $k_3=1.7$ and $V_o = 0.1$. Points are shown for 2000 randomized initial conditions with randomized initial field densities.}%each of the three fields would yield in the absence of the other two. }
\end{figure}

The next step in the analysis is to understand the effect of super-inflationary phase on the multiple field system. In particular, an important question is whether the number of e-foldings during super-inflation can further assist the assisted inflation scenario. We investigated the number of e-foldings allowed during super-inflation for potentials with various parameters in the case of a 3 fields.  An example of the results for steep potentials $V_i$ is shown in Fig. 13, for $k_1=1.5$, $k_2=1.6$, $k_3=1.7$ and $V_o = 0.1$, where 2000 data points for number of e-foldings obtained by simulations with varying initial conditions are plotted. These were obtained as follows. First a pseudo-random initial density, $\rho_1$, is chosen between $0.001\rmax$ and $0.99 \rmax$. Then $\rho_2$ is chosen in a pseudo-random way with value between 0 and $\rmax - \rho_1$. Similarly, the energy density $\rho_3$ is chosen. The densities are then permuted pseudo-randomly, to prevent $\rho_1$ from being larger on average than the others. These are then paired with the fields $\phi, \phi_2, \phi_3$, and a  large array of initial conditions is generated along the density level curve defined by $\rho =\rho_i$ (where $i = 1,2,3$) on the portion of this curve nearest to the potential regime. This placement, along with the fact that the initial density generation method statistically favors high initial $\rho_{total} = \sum \rho_i$, and helps produce bounces with total equation of state $w \not\approx 1$. Each field is separately assigned an initial condition from the appropriate array generated in the previous step, and the LQC equations are solved with these initial conditions. These solutions are used to find the number of e-foldings during super-inflation, and the process is repeated. The effective steepness parameter for the above case is $k_{\rm{eq}} \approx 0.92$. Comparing our result with the single field case (see Fig. 10), we find that for all initial conditions, the number of e-foldings during super-inflation is less than the maximum number of e-foldings for the single field case with the steepness parameter $k = k_{\rm eq}$. %Thus, super-inflation does not significantly enhance the number of e-foldings for assisted inflation for steepness parameters $k_i > \sqrt{2}$.

\begin{figure}[tbh!]
%\begin{center}
%\label{multifield2}
\includegraphics[angle=0,width=0.45\textwidth]{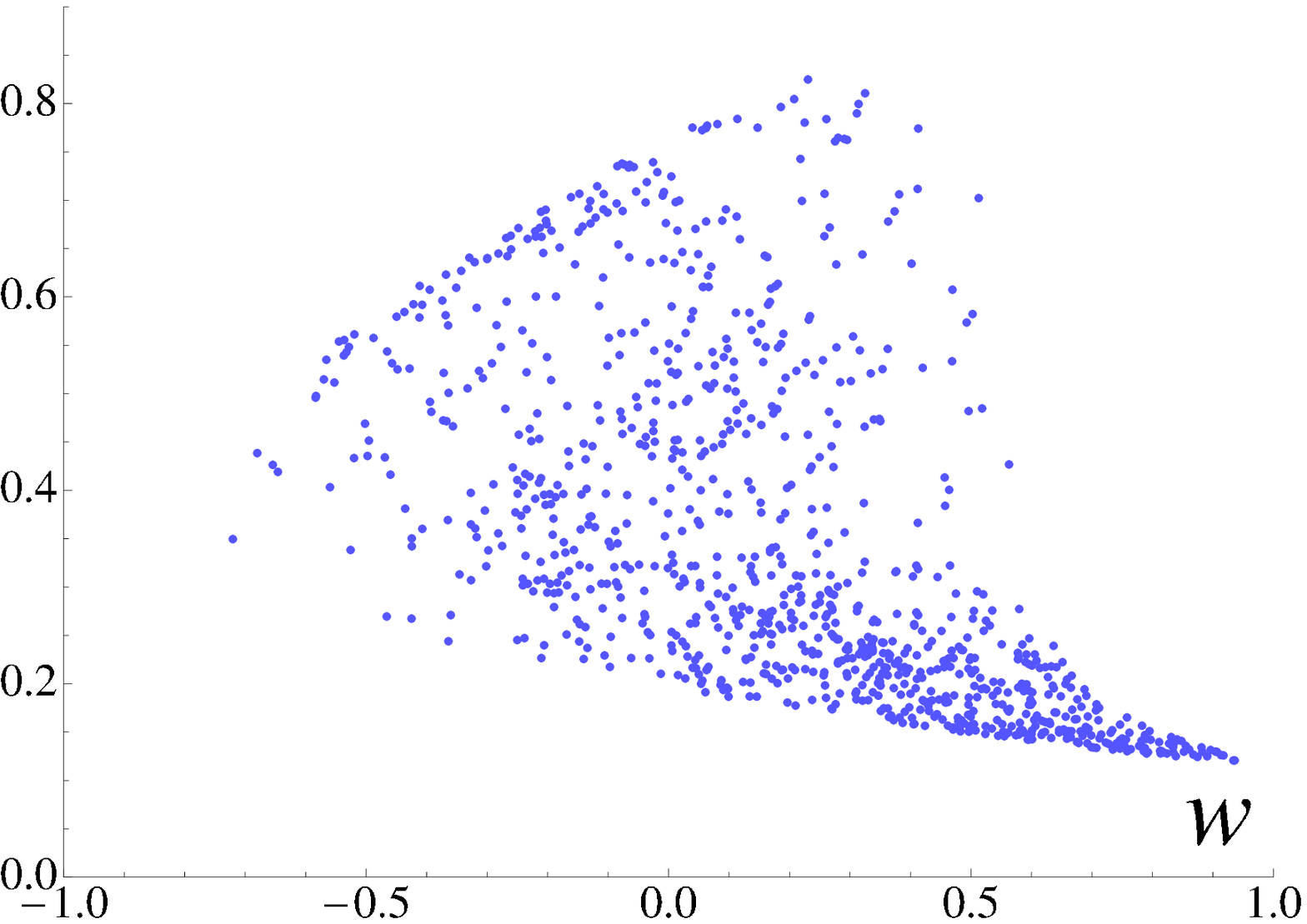}
\includegraphics[angle=0,width=0.45\textwidth]{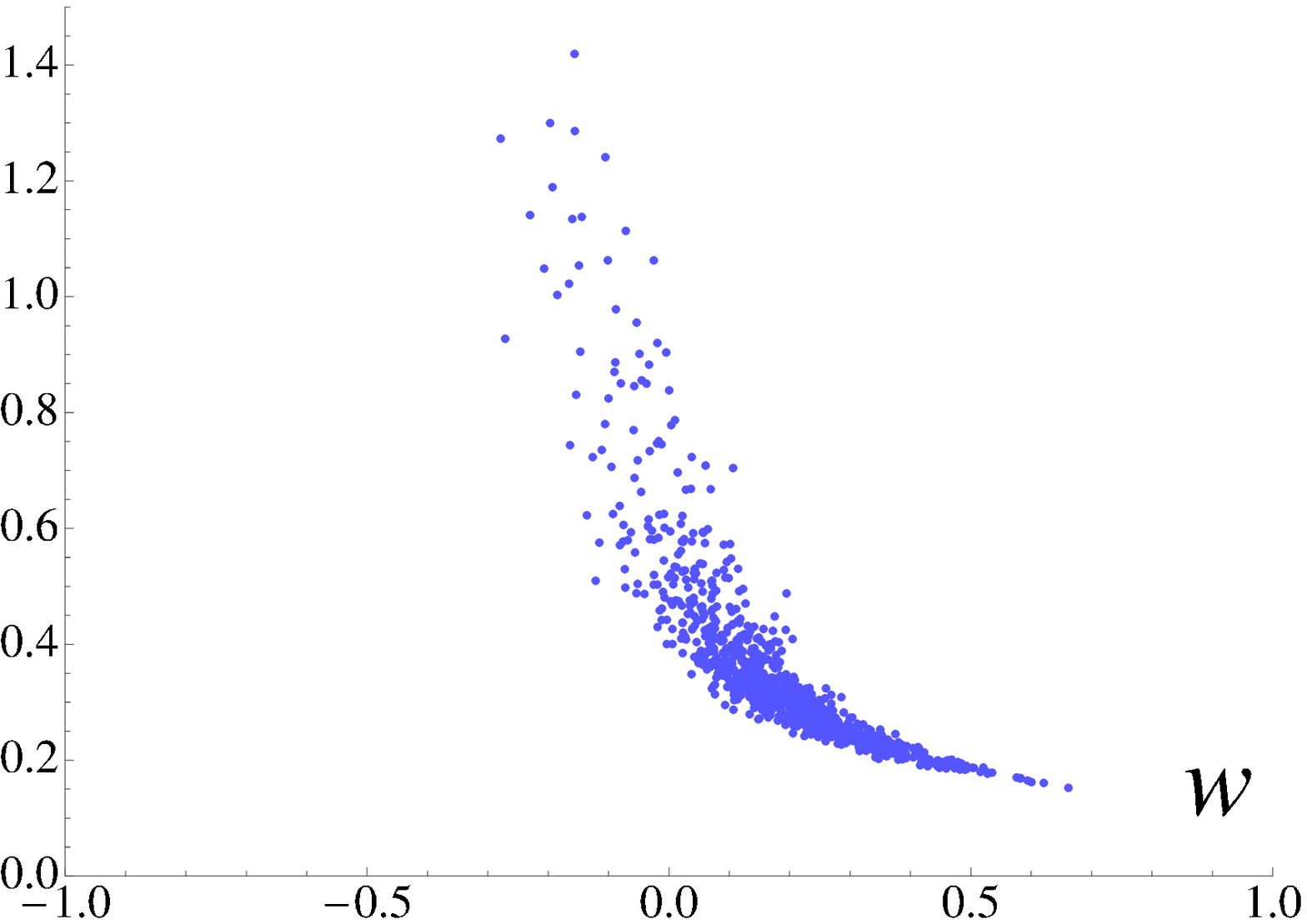}
\caption{The left plot shows the number of e-foldings obtained during super-inflation for three fields with $k_i = 1.5$. The right plot corresponds to 20 fields with the same $k_i$'s. In each plot 1000 initial conditions are considered, randomized as in Fig. 13 with the added constraint that the total initial density of each field is $0.9 \rmax$, divided by the number of fields. }
%}%each of the three fields would yield in the absence of the other two. }
\end{figure}

For the three field scenario, though the number of e-foldings during super-inflation are not significant, they increase on addition of more fields. This is confirmed from the plots in Fig. 14 where we have contrasted the three field system with 20 fields, each having same steepness parameter $k_i = 1.5$. The initial density in both plots corresponds to $0.9 \rmax$, evenly distributed amongst fields and initial conditions randomized as above. Notable differences arise on changing the number of fields. First, the plots show that the amount of e-foldings during super-inflation can increase with an increase in the number of fields in the assisted inflation scenario in LQC.
  Second, we find that in the case of three fields, the number of e-foldings during super-inflation are closer to the value in the equivalent single field case (with $k_{\rm{eq}} = 0.866$), than for the 20 fields (with $k_{\rm{eq}} = 0.335$). Further, the distribution of points in the plot of 20 fields is in narrow region than in the case of three fields. Our simulations showed that this is due to lower initial energy densities of the individual fields for the case of 20 fields.

Finally, let us understand the way the number of e-foldings during super-inflation change if one chooses a shallower potential.
To examine this we performed various simulations involving $k_i < \sqrt{2}$ for three fields. An example of one of the results is plotted in the left figure of Fig. 14. Here the
steepness parameters are equal for all the fields, $k_1 = k_2 = k_3 = 0.3984$ with $k_{\rm{eq}} = 0.23$. Following the process of providing initial conditions as for Fig. 13, e-foldings in the super-inflationary phase were obtained for 2500 numerical runs.
We found that almost all points lie in the area defined by two curves for the single field cases, one with steepness parameter $k =k_{\rm{eq}}= 0.23$ and another with $k' =k_i= 0.3984$. Further, data points display a tendency to clump near the boundaries of the limiting $k$ and $k'$ curves in some regions. This plot is to be contrasted with the second plot in Fig. 14, obtained with different values of $k_1, k_2$ and $k_3$ yielding  $k_{\rm{eq}} = 0.23$. In this case the data points are obtained as follows. Note that the power in the asymptotic scale factor solution can be written as $n_{\rm{eq}}=\sum n_i$. This power can be randomized in the same way as the randomization of densities in Fig. 13.
First,  $n_1$ is chosen pseudo-randomly between $0.001n_{\rm{eq}}$ and $0.99n_{\rm{eq}}$. Then $n_2$ is  chosen pseudo-randomly between $0$ and $n_{\rm{eq}}- n_1$. Similarly, $n_3$ is chosen such that  $\sum n_i = n_{\rm{eq}}$. We then use the relation $k_i=\sqrt{2/n_i}$ to determine the steepness of each field. Then the initial conditions are randomized as in Figure 13. This process was repeated 2500 times. We see that data points are now found outside their previous boundaries in left plot in Fig. 14, on the bottom right and left, and are uniformly distributed in comparison. This behavior shows that the distribution of points in the plot of number of e-foldings versus equation of state is sensitive to the individual steepness of different potentials in assisted inflation. However, note that no points were found which led to more e-foldings than the maximum number of e-foldings during super-inflation for the single field case. One can thus conclude that maximum number of e-foldings in the single field case achieved during super-inflation provides a good upper bound on the number of e-foldings during super-inflation for the multi-field scenarios with exponential potential.

\begin{figure}[tbh!]
%\begin{center}
%\label{multifield2}
\includegraphics[angle=0,width=0.45\textwidth]{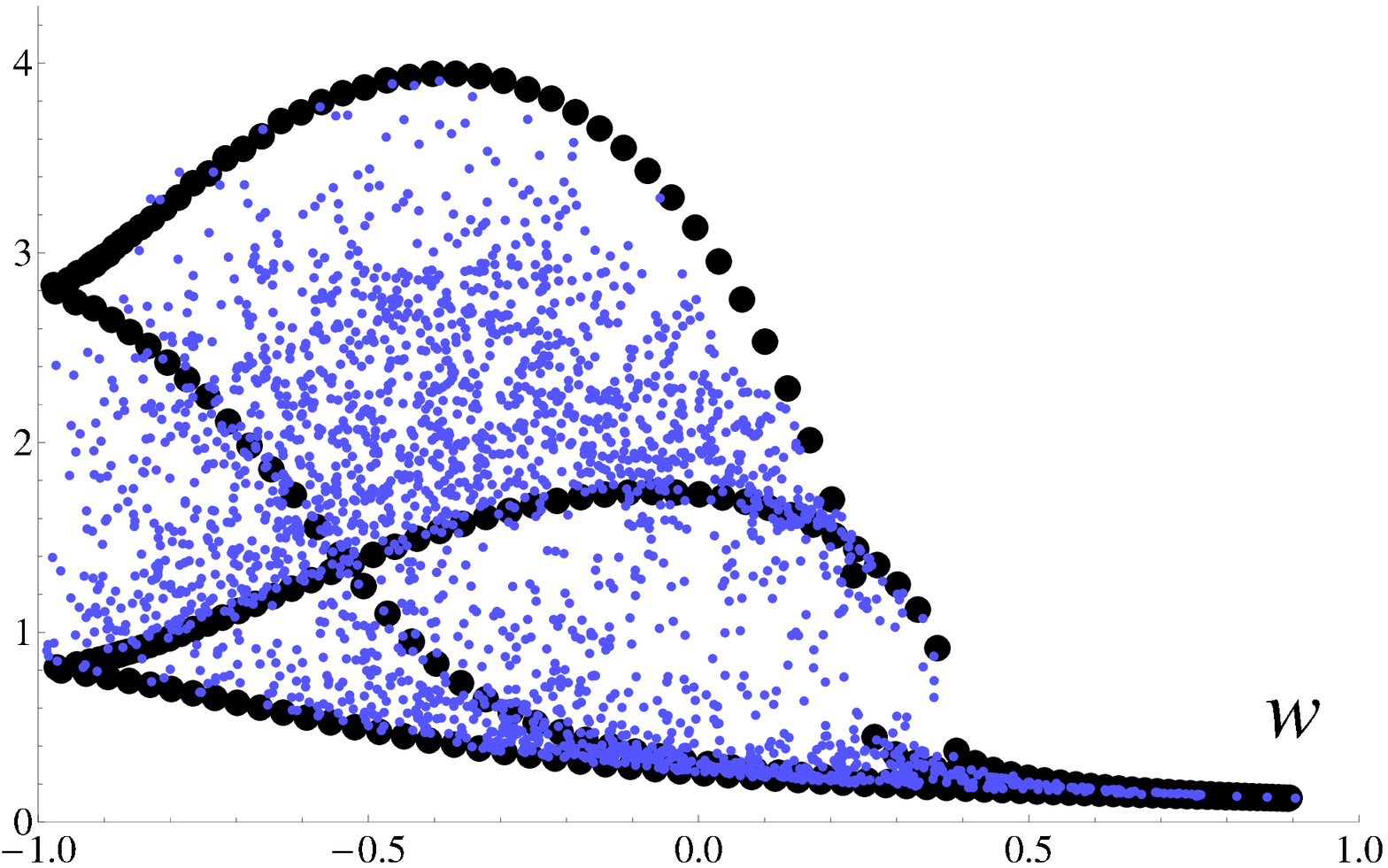}
\includegraphics[angle=0,width=0.45\textwidth]{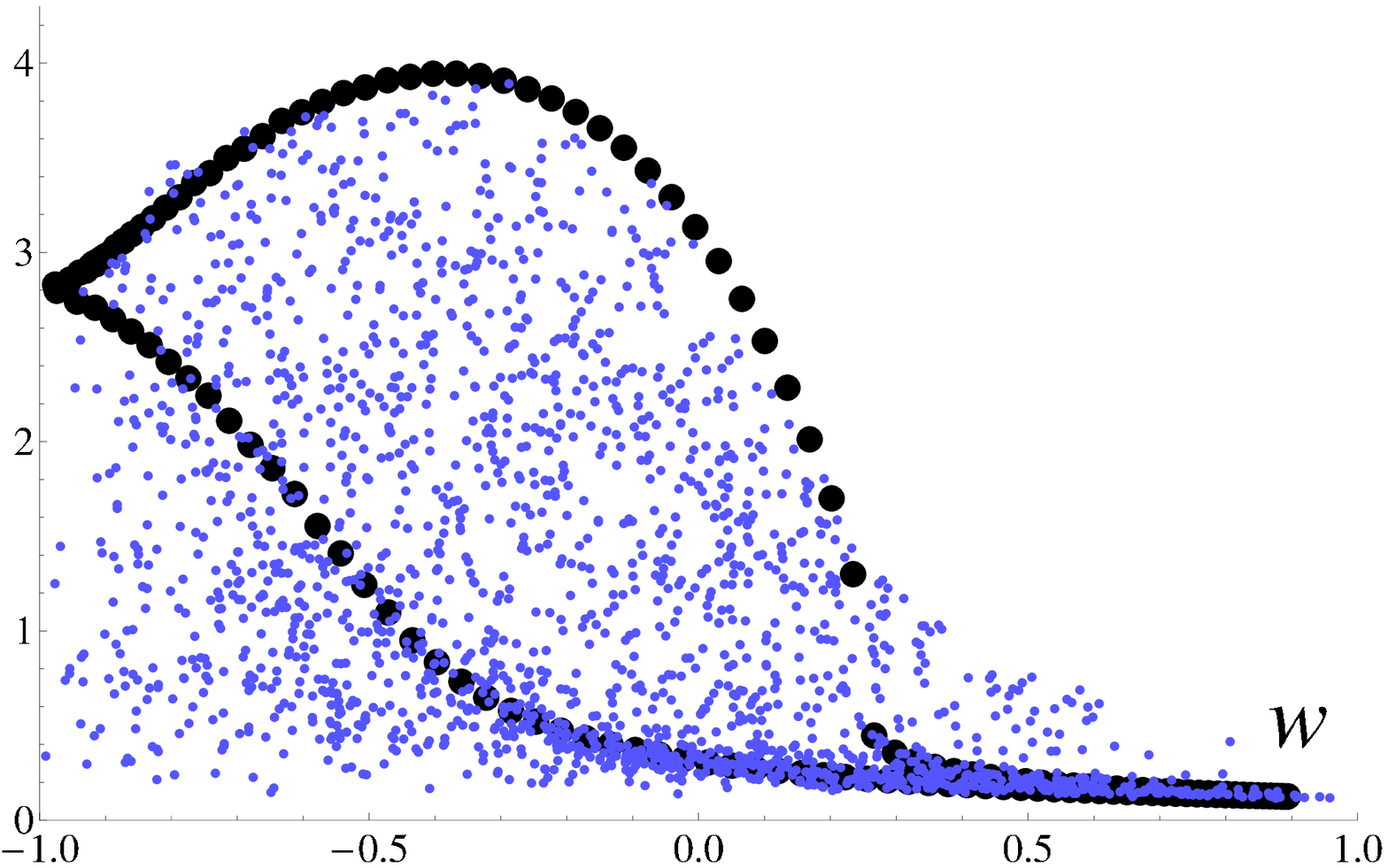}
\caption{Left plot shows the number of e-foldings during super-inflation for three field system with $k_1=k_2=k_3=0.3984$ and $V_o = 0.1$. The thick dotted upper curve corresponds to the e-foldings during super-inflation for the single field case for $k= k_{\rm{eq}} = 0.23$ and the lower one for $k' =k_i= 0.3984$. The right plot shows the number of e-foldings during super-inflation for the three field system for $V_o = 0.1$ and $k_{\rm{eq}} = 0.23$ with differing $k_i's$ and initial conditions chosen in a pseudo-random way as described in the text. }
\end{figure}

\section{Summary}
Scalar fields with exponential potentials appear naturally in various theories of gravity at high energies. These potentials have been used to construct models of power-law and assisted inflation. However, as in the case of the chaotic $\phi^2$ inflation model, the past evolution in these models is generically singular in GR. It has been hoped that incorporation of quantum gravity effects would resolve the initial singularity in these models and provide insights on the new physics. In this work, we probe this issue within the framework of LQC, which has been successful in resolving various cosmological singularities \cite{as}. Using the effective spacetime description, we analyze the dynamical evolution of single and multiple fields with positive exponential self-interaction potentials. We show that generically non-singular power-law and assisted inflation scenarios can be constructed in LQC. Unlike the classical theory, where all solutions are singular, past evolution using modified dynamical equations in LQC generically result in a bounce of the scale factor when energy density becomes equal to $\rmax \approx 0.41 \rho_{\rm{Pl}}$. As an important element of new physics at the Planck scale, a phase of super-inflation which occurs when $\rmax > \rho > \rmax/2$ precedes the classical inflationary phase. This phase  contributes additional e-foldings of inflation to those in the classical inflationary phase, which can potentially affect the viability of these models by modifying the constraints on the steepness parameter in power-law and assisted inflation scenario. The number of e-foldings contributed by super-inflation can be large for non-kinetic  dominated bounces. Our results show that as the steepness parameter $k$ (or $k_{\rm{eq}}$ in assisted inflation) decreases, the number of possible e-foldings which can occur during super-inflation increases. In assisted inflation, this can be achieved by adding more fields without changing the steepness of the field potentials.
This result indicates that if the exponential potential is suitably modified to end the inflation,  a given number of e-foldings can be achieved in LQC in principle with a lesser number of fields than in GR. Due to above reasons, quantum gravity effects as understood in LQC, can potentially further assist assisted inflation scenario.

In the power-law and assisted inflation scenarios, at late times in the expanding branch, classical solutions approach an attractor, and at early times, in the backward evolution of the expanding branch,
dynamical trajectories asymptotically approach those of the massless scalar field near the initial singularity.
 Dynamical trajectories in the contracting branch, disjoint from the expanding branch in the classical theory, in the backward evolution approach the attractor at early times. We find that in LQC, due to generic resolution of the big bang singularity, the situation changes dramatically. In the forward evolution of the expanding branch, dynamical trajectories in LQC approach the attractor of the classical theory. However, we find that before that occurs, dynamical trajectories approach a non-classical curve in the phase space for $\rho \sim \rmax/2$, which approaches  the classical attractor as the energy density decreases. Existence of this curve is also found in the pre-bounce regime, in the backward evolution of the contracting branch, which approaches the backward time attractor at early times. %In our numerical simulations, we did not find LQC attractors to exist in the regime $\rho \sim \rmax$.
Thus, quantum geometry effects in LQC  bridge the pre-bounce repeller with the post-bounce attractor in the forward time evolution.

In the classical theory dynamical trajectories asymptotically approach those of the massless scalar field near the initial singularity, and the approach to singularity is thus dominated by kinetic part of the matter Hamiltonian.  In LQC, since all solutions bounce at a non-zero volume, solutions at the bounce can be dominated by kinetic as well as potential terms. Detailed analysis of the phase space trajectories reveals that the equation of state at the bounce varies from $w = 1$ to $w = -1$ depending on the initial conditions and the steepness of the exponential potential. The maximum number of e-foldings during super-inflation increases as the steepness parameter decreases. The dependence is non-linear, and the number of e-foldings can be very large for shallow potentials.  An interesting feature of this model is that the maximum number of e-foldings during super-inflation, depending on steepness parameter, occur for different values of equation of state at the bounce. As the steepness parameter decreases, the equation of state at the bounce which yields maximum number of e-folding during super-inflation shifts towards $w \approx -1$. For steeper potentials, the equation of state at the bounce yielding maximum number of e-foldings shifts towards $w \approx 1$. %This behavior contrasts with that in the $m ^2 \phi^2$ potential, where $w \approx -1$ at the bounce leads to maximum number of e-foldings during super-inflation for arbitrary values of the parameter $m$.
Analysis of the multi-field system in assisted inflation scenario shows that the number of e-foldings obtained during super-inflation are roughly bounded above by the single field case with an equivalent steepness parameter. We find that the likelihood of obtaining high e-foldings during super-inflation for a given set of initial conditions with same equivalent steepness parameter changes if the individual steepness parameters of multiple fields change. The number of e-foldings during super-inflation can increase with an increase in the number of fields in assisted inflation.
%Thus, if a specified number of e-foldings are required in  assisted inflation, they can be obtained by a lesser number of fields than in GR.

An important question is the way above results are affected if we go beyond the homogeneity and isotropy assumption in loop quantum cosmology. The first step in this direction would be to analyze the dynamics with exponential potentials in anisotropic situations, such as in the Bianchi-I model. Due to availability of the underlying loop quantization \cite{awe2}, which also indicates a generic resolution of strong singularities at the effective spacetime  level \cite{ps11}, it is expected that the classical singularity will be avoided in a qualitatively similar manner as in the isotropic model. However, due to non-zero anisotropic shear, the resulting physics at the Planck scale is expected to be richer.  As one of the features, the approach to the classical singularity would be affected by the presence of anisotropies. Unlike the isotropic model, where the singularities in the classical theory are point-like, the approach to singularity may be barrel, cigar or pancake type in anisotropic models. This could also potentially affect the pre-bounce attractor behavior. Further, in the case of anisotropic models, the phase of super-inflation becomes more non-trivial due to the bounces in three different scale factors. It will be thus important to understand in future work the role of anisotropies on e-foldings obtained in the bounce regime of the power-law and assisted inflation scenarios in Bianchi models. The next step in understanding above results in a broader setting would require going beyond the homogeneity assumption. At the current stage of research, loop quantum effects as discussed in this manuscript are yet to be derived from loop quantum gravity and the way inhomogeneities affect the background dynamics is also not understood. However, important insights on connection of loop quantum cosmology and loop quantum gravity \cite{engle,tim}, and inclusion of inhomogeneities using spinfoam techniques have been gained \cite{carlo,livine}. Though it leaves the discussion of affects of inhomogeneities on above results open, if the singularity resolution in loop quantum gravity occurs via a bounce, we expect some of the main results to remain unaffected, at least qualitatively. This expectation is based on the observation that a bounce of the homogeneous part of the metric would lead to a phase of super-inflation, which under the assumption that inhomogeneities remain small and dynamics remaining stable, would cause similar effects on power-law and assisted inflation scenarios. Whether or not this expectation holds, will be answered by future research on inclusion of inhomogeneities in this setting.

Results obtained in this work enable us to rigorously understand the dynamics for exponential potentials in the presence of single and multiple fields, and bring forward some effects of the super-inflationary phase which were not noticed earlier with other forms of matter and potentials.  In future work, it will be important to understand the effect that this phase and the increase in the number of e-foldings have on observational constraints on power-law and assisted inflation scenarios. The power-law inflation was recently briefly considered in the context of LQC \cite{mb}, however one would need to go beyond the underlying approximations in Ref. \cite{mb} to study in detail the implications of our results on the cosmological perturbations. It will be also interesting to extend this analysis by including different fluids, and study the effects of the scalar-fluid interaction on the dependence of the number of e-foldings during super-inflation on the steepness parameter and the equation of state. Finally, it will be instructive to analytically understand the approach of dynamical trajectories in LQC, particularly in the Planck regime, to the inflationary  attractor in the future evolution.

\vskip0.5cm

\acknowledgments
We thank Brajesh Gupt for extensive discussions and Jorge Pullin for useful comments. E.R. thanks the REU program in Physics and Astronomy at LSU during which a large part of this work was completed. This work is supported by NSF grants PHY1068743 and PHY1004822.

\end{document}